\documentclass[useAMS,usenatbib]{mn2e_mod}
\usepackage{amssymb,hyperref,subfigure,epsfig,multirow}

\newcommand{\Msun}{M_{\odot}}
\newcommand{\Mjup}{M_{\mathrm{Jup}}}
\newcommand{\Mearth}{M_{\oplus}}
\newcommand{\Table}[1]{Table~\ref{#1}}
\newcommand{\Figure}[1]{Figure~\ref{#1}}

\newcommand{\rev}[1]{#1}

\title[Traditional formation scenarios and the 4:3 resonance]{Traditional formation scenarios fail to explain 4:3 mean motion resonances}
\author[H.~Rein,\,M.~J.~Payne,\,D.~Veras\,\&\,E.~B.~Ford]{Hanno~Rein$^{1}$\footnotemark[1],\,Matthew~J.~Payne$^{2}$,\,Dimitri Veras$^{3}$\,\&\,Eric~B.~Ford$^{2}$
 \\
  $^1$Institute for Advanced Study, 1 Einstein Drive, Princeton, NJ 08540, USA\\
  $^2$Department of Astronomy, University of Florida, 211 Bryant Space Science Center, PO Box 112055 Gainesville, FL, 32611-2055, USA\\
  $^3$Institute of Astronomy, University of Cambridge, Madingley Road, Cambridge CB3 0HA, UK\\}
\date{Submitted: \today.}

\begin{document}
\maketitle
\begin{abstract}
At least two multi-planetary systems in a 4:3~mean motion resonance have been found by radial velocity surveys\footnotemark[2].
These planets are gas giants and \rev{the systems are} only stable when protected by a resonance. 
Additionally the Kepler mission has detected at least 4 strong candidate planetary systems with a period ratio close to~4:3.

This paper investigates traditional dynamical scenarios for the formation of these systems.
We systematically study migration scenarios with both $N$-body and hydrodynamic simulations.
We investigate scenarios involving the in-situ formation of two planets in resonance.
We look at the results from finely tuned planet-planet scattering simulations with gas disk damping.
Finally, we investigate a formation scenario involving isolation-mass embryos.

Although the combined planet-planet scattering and damping scenario seems promising, none of the above scenarios is successful in forming enough systems in 4:3 resonance with planetary masses similar to the observed ones.
This is a negative result but it has important implications for planet formation.
Previous studies were successful in forming 2:1 and 3:2 resonances.
This is generally believed to be evidence of planet migration.
We highlight the main differences between those studies and our failure in forming a 4:3 resonance. 
We also speculate on more exotic and complicated ideas.
These results will guide future investigators toward exploring the above scenarios and alternative mechanisms in a more general framework.

\vspace{0.5cm}
\end{abstract}

\label{firstpage}
\footnotetext[1]{e-mail: \rm{\url{rein@ias.edu}.}}
\footnotetext[2]{The discovery paper announcing the second system is currently in preparation (Giguere et al., in prep).}

\section{Introduction}\label{sec:introduction}

To date, 777\footnote{See e.g. \url{http://exoplanetapp.com}.} extra-solar planets have been discovered via numerous detection techniques, including  
pulsar timing \citep[e.g.][]{Wol1992}, 
radial velocity \citep[RV, see e.g.][]{Mayor1995},  
Transits \citep[e.g.][]{Charbonneau2000}, 
and micro-lensing \citep[e.g.][]{microlensing}, 
while thousands of candidate systems from the Kepler transit mission await confirmation \citep{Batalha2012}.

A significant fraction ($\sim13\%$) of the known planetary systems have been confirmed to possess systems of multiple planets.
Multi-planet systems can provide valuable information on their history that single planet systems cannot.

For example, multiple highly excited planets may indicate early dynamical instabilities \citep[][]{Rasio1996}.
The existence of mean motion resonances (MMRs) on the other hand suggests that convergent migration occurred in the presence of dissipative forces.
Numerous examples of resonant systems are known, both in extra-solar planetary systems and solar system satellites such as the 1:2:4 Laplace resonance in the Io-Europa-Ganymede system.
The most studied planetary system in a MMR is Gliese~876 \citep[e.g.][]{MarcyButler01,LeePeale01,LeePeale2002,SnellgrovePapaloizouNelson01,NelsonPapaloizou2002,beamic2003,veras2007,ReinPapaloizou2010}.
All these studies suggest that migration (albeit potentially driven by a variety of mechanisms) is an important factor in the sculpting of planetary systems.

Convergent migration into closely spaced resonances (e.g. 3:2) has been shown to be plausible, but difficult \citep[][]{ReinPapaloizouKley2010}. 
The difficulties arise because planets that initially form far apart need to migrate through more widely spaced resonances (e.g. the 2:1 MMR) before subsequently being captured into the more closely spaced (e.g. 3:2) MMR. 
To avoid the requirement for fine-tuning of initial locations, this requires relatively high migration rates \citep{ReinPapaloizouKley2010}.
Even more closely spaced 4:3 resonances are suspected for some systems such as HD~200964 \citep{JohnsonPayne2011}, and KOI~115 \citep{Borucki2011}, suggesting that even more violent migration histories must have occurred to allow such systems to skip through the exterior 2:1 and 3:2~resonances before going on to be captured into the observed 4:3~resonance.

\rev{We investigate the formation of 4:3~resonances in massive systems, i.e. planets with masses up to several times that of Jupiter.} 
We examine mechanisms which include smooth migration, in-situ formation, scattering and damping.
We demonstrate that the simple smooth migration mechanisms suspected to form the known 2:1 and 3:2 systems cannot plausibly form systems of massive planets in 4:3 resonances. 

\rev{We also study the formation of lower mass planets and find that these can readily form from a series of isolation-mass embryos in tightly-packed systems which then lead to 4:3 (and even closer) resonances.}
But these mechanisms underestimate the number of tightly packed systems in close resonances with multiple massive planets.
The rate of detection of these multi-planetary systems (period ratios of 4:3 and closer) can therefore tell us valuable, not directly observable information about the formation history of extra-solar planetary systems and the solar system itself. 

In Section~\ref{sec:observations} we summarize observational results of closely packed planetary systems.
Then we discuss the phase space and stability of massive systems in Section~\ref{sec:stability}.
The results of many different formation scenarios are presented in Section~\ref{sec:formation}. 
This is the main part of our paper.
\rev{We then extend the study to lower mass systems in 4:3 resonance  and show that there is no difficulty in forming these in Section~\ref{sec:smallmass}.}
Finally we summarize and discuss these results in Section~\ref{sec:conclusions}.

\section{Observational evidence for 4:3 resonances}\label{sec:observations}
As noted in Section~\ref{sec:introduction}, while many systems are thought to be in (or to be close to) a variety of mean-motion resonances, the population of systems suspected to occupy the closely spaced 4:3 MMR is much smaller. 
In this section we review the systems, both solar system satellites as well as exoplanets, which are suspected of populating 4:3 resonances.

\subsection{4:3 resonances in the Solar System}\label{sec:observations:solarsystem}
\rev{The best known pair} of solar system satellites or planets which are locked in a 4:3~MMR with each other are Titan and Hyperion.  
With a mass of $5.6 \cdot 10^{18}$~kg \citep{thoetal2007}, Hyperion is 168 times less massive than Ceres and almost five orders of magnitude less massive than Titan, meaning that Hyperion effectively acts as a test particle and has a negligible effect on Titan's orbit.
Hence, this resonance is dominated by a single term in the disturbing function (containing the longitude of pericenter of Hyperion and the mean longitudes of both satellites).
 
The satellites' motion has been well-characterized by observations over several decades \citep{taylor1984} and the stability afforded by the resonance has been considered in great detail \citep{coletal1974,bevetal1980,stellmacher1999}.
Although recently \cite{beaetal2006} suggested that Titan and Hyperion is an example of a system captured into resonance due to migration, \cite{bevetal1980} came to a different conclusion.
The latter suggested that Hyperion was formed at its present location and cast doubt on a scenario where Titan and Hyperion achieved their current configuration through smooth differential tidal evolution across the chaotic zone.
They further argued that a possible reason why Hyperion was accreted together in the 4:3~libration zone instead of the 3:2~libration zone is because in the former, more restricted region, the relative velocities of surviving planetesimals were small enough for coagulation to occur.

\rev{Other cases of 4:3 resonances in the solar system include the asteroid Thule which is in the 4:3 resonance with Jupiter (for more objects see page 49 in \citealt{SolarSystemBeyondNeptune2008}).}
 
Unfortunately, \rev{none of the above considerations translate} easily into a 4:3~configuration with two massive planets.
In this case, multiple terms in the disturbing function must be considered and stable libration zones are not as well characterized.
However, as demonstrated later, capture into this resonance due to migration still proves to be difficult.

\subsection{4:3 resonances in exo-planetary systems}\label{sec:observations:exo}

\subsubsection{Radial velocity systems}\label{sec:observations:exo:rv}

\begin{figure}
\centering
\psfig{figure=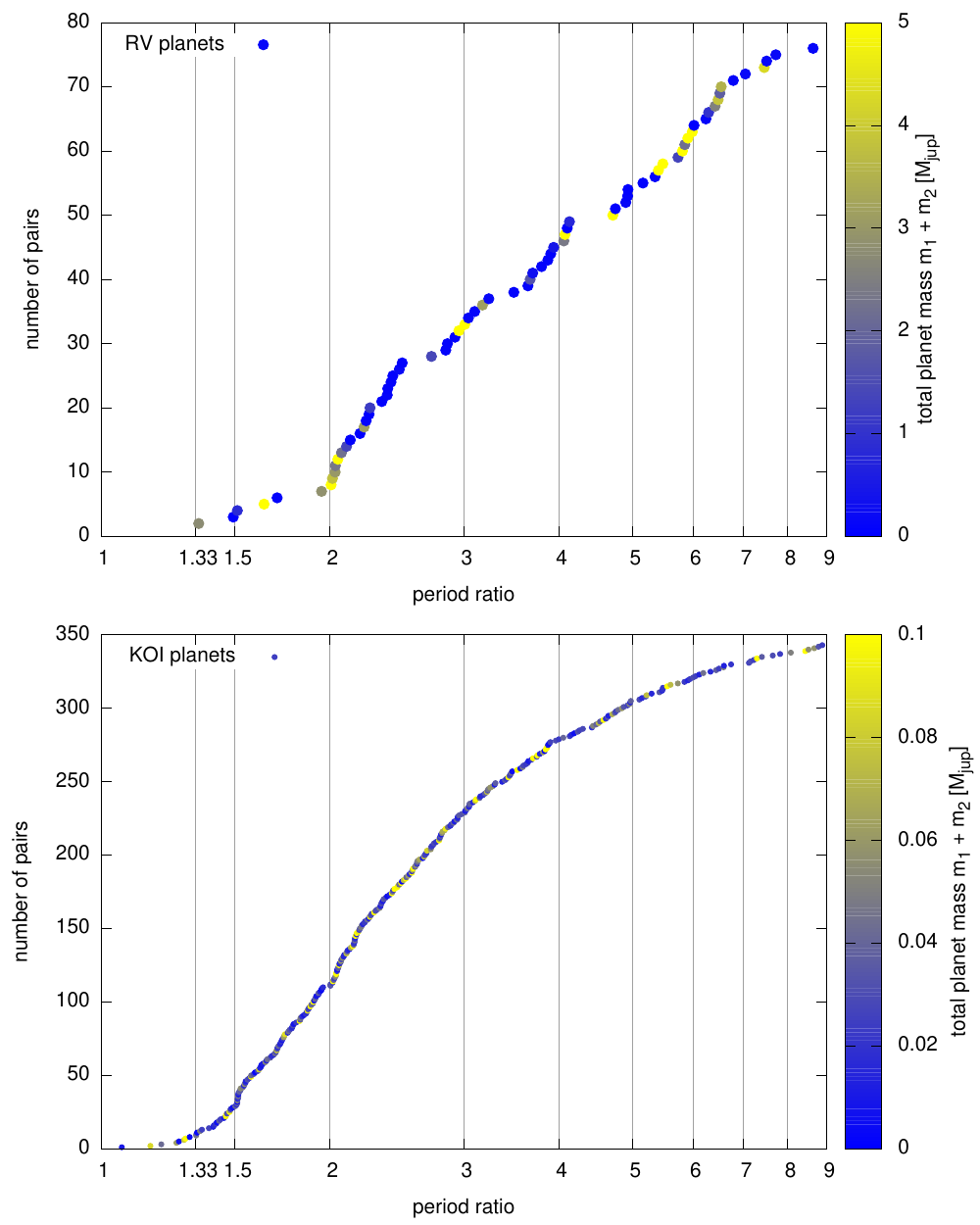,angle=0,width=\columnwidth}
\caption{Cumulative distribution function of all multi-planetary systems discovered with the radial velocity method (top) and all Kepler candidates (KOIs, bottom). 
The color indicates the total mass of the planet pair, $m_1+m_2$. }
\label{fig:cdf}
\end{figure}

\begin{table}
\centering
\begin{tabular}{l |c | l | cc| cc | c} \hline
Star & $M_* [\Msun]$  & Planet           & $m [\Mjup]$   & $P$  [days]       &$P_2/P_1$     \\
\hline
\hline
\multirow{2}{*}{HD200964} & \multirow{2}{*}{1.44} & b & 1.99 & 613.8 & \multirow{2}{*}{1.34}  \\
 &     &  c    & 0.90    & 825.0     &  \\ 
\hline \end{tabular}
\caption{\label{tab:rv} \rev{Detail of the planetary system HD200964 which has been detected by radial velocity variations  and is in or near a 4:3 mean motion resonance. A second, very similar system has been detected but has not been announced by the time this paper was submitted (Giguere et al., in prep).}}
\end{table}

There are currently two multi-planetary system discovered by the radial velocity method which are \rev{reported to be} in or near a 4:3 mean motion resonance. 

HD~200964 consists of two planets with masses $m_1 = 1.8 \Mjup$ and $m_2=0.9 \Mjup$ \citep{JohnsonPayne2011}.
The period ratio is close to $1.33$ which suggests that the system is in a 4:3~mean motion resonance.
The results of a Monte Carlo fitting routine that includes a penalty for unstable systems strongly favor systems in resonance \citep[][see also Section~\ref{sec:stability}]{JohnsonPayne2011}. 
Another planetary system is suspected to also be in or near a 4:3 MMR (Giguere et al., in prep, private communication). 
Both systems consist of two massive (and most likely gaseous) planets on wide (a few AU) orbits.
Both systems are subject to a short dynamical instability if the planets are not protected by a mean motion resonance. 
Their parameters are listed in \Table{tab:rv}.

In the top panel of Figure~\ref{fig:cdf} we plot the cumulative distribution function of all the multi-planetary systems that have been discovered with the radial velocity method.
In systems with more than two planets, each pair is treated independently.
To compile this data set, we made use of the Open Exoplanet Catalogue\footnote{\url{https://github.com/hannorein/open\_exoplanet\_catalogue}}.
One can clearly see the tendency for planets to pile up near integer ratios of the period ratio such as~2:1~and~3:1. 
This is usually attributed to resonant capture during the migration phase \citep{LeePeale2002}.
The number of planets near the 4:3 resonance is far too small to make a statistical argument at this time. 
We also color code the total mass of the two planets. 
Note that mostly high mass planets get captured in the 2:1 and 3:2 resonances, whereas lower mass planets get preferentially captured into more closely spaced resonances.

In this paper, we investigate the conditions under which systems on the far left side of this plot form.

\rev{
The reported best fit solutions of HD~200964 system puts it in a 4:3 resonance. 
However, there are other orbital solutions which are stable and cannot be ruled out with high confidence. 
For example, HD~200964 could also be in a 3:2 resonance \citep{JohnsonPayne2011}.
With the currently published RV data-points, this results in a higher $\chi^2$ value, therefore not being the best, but still a possible fit.
This paper is concerned about the formation scenarios of the report systems.
Fitting radial velocity data is a notoriously difficult job. 
We do not attempt to redo the analysis of \cite{JohnsonPayne2011}.
}

\rev{
As we will show below, it is very difficult to get the system into the 4:3 resonance without fine-tuded initial conditions. 
One could therefore take our results and use it as a strong prior while fitting the RV light-curve, rejecting systems in 4:3 resonance. 
We do not want to go that far and think this is actually dangerous.
If there is a new formation mechanism that we did not take into account, one can easily draw a wrong conclusion. 
}

\subsubsection{Kepler systems }\label{sec:observations:exo:kepler}

\begin{table}
\begin{tabular}{l | cc| cc | c}
\hline
KOI & $R_1 [R_{\oplus}]$             & $R_2 [R_{\oplus}]$             & $P_1$  [days]       & $P_2$ [days]     &$P_2/P_1$     \\
\hline
\hline
115 & 3.4    & 2.2    & 5.41    & 7.13    & 1.32 \\
543 & 1.5    & 1.9    & 3.14    & 4.3     & 1.37 \\ 
749 & 1.4    & 2.0    & 3.94    & 5.35    & 1.36 \\
787 & 2.9    & 2.2    & 4.43    & 5.69    & 1.28 \\
\hline
\end{tabular}
\caption{\label{tab:KOI} Kepler candidate systems which may occupy a 4:3~mean motion resonance. 
}
\end{table}

In \Table{tab:KOI} we list the four Kepler candidate systems which may occupy a~4:3~MMR \citep{Batalha2012}. 
The first column lists the Kepler Object of Interest Number of the candidate system. 
The second and third columns list the planet radius in units of Earth-radii. 
The fourth and fifth columns list the orbital periods in days. 
The last column lists the ratio of the periods. 
We follow the procedure outlined in \citep{verfor2011} and exclude any systems which are unlikely to be in resonance despite having period ratios close to~1.33. 
Note that all of the Kepler systems listed in \Table{tab:KOI} are smaller and closer-in than the RV systems listed in \Table{tab:rv}: the largest of the KOIs has a radius less than that of Uranus, and all have orbital periods less than 8 days.
It is thus not unreasonable to assume that their formation mechanism differs significantly from that of the massive planets at larger semi-major axes.

We plot the cumulative distribution of all Kepler planet candidates \citep[KOIs,][]{Batalha2012} in the bottom panel of Figure~\ref{fig:cdf}.
As in Section~\ref{sec:observations:exo:rv}, for systems with more than two planets, every pair in that system is treated independently.
Although there are clear features in this distribution near the 3:2 and 2:1 resonances, no statistically significant accumulation of planets can be seen near the 4:3 resonance.
Furthermore, in contrast to the RV systems, there is no strong correlation between mass and the proximity to a resonance.

\rev{It is worth pointing out the Kepler systems Kepler~36 and KOI~262.
These systems are near the 6:7~MMR and 5:6 MMR, respectively \citep{Carter2012,Fabrycky2012}, i.e. even closer spaced than the 4:3~MMR that we study here.
The masses of those planets are all below nine Earth masses.
}

\section{Phase space and stability analysis}\label{sec:stability}
\subsection{Expansion of the disturbing function}
\begin{figure*}
\centering
\subfigure[Planet masses  $m_1=1.8\,\Mjup$, $m_2=0.9\,\Mjup$. The outer planet's orbit is $a_2=3.0$~AU with~$e_2=0.01$.]{
\psfig{figure=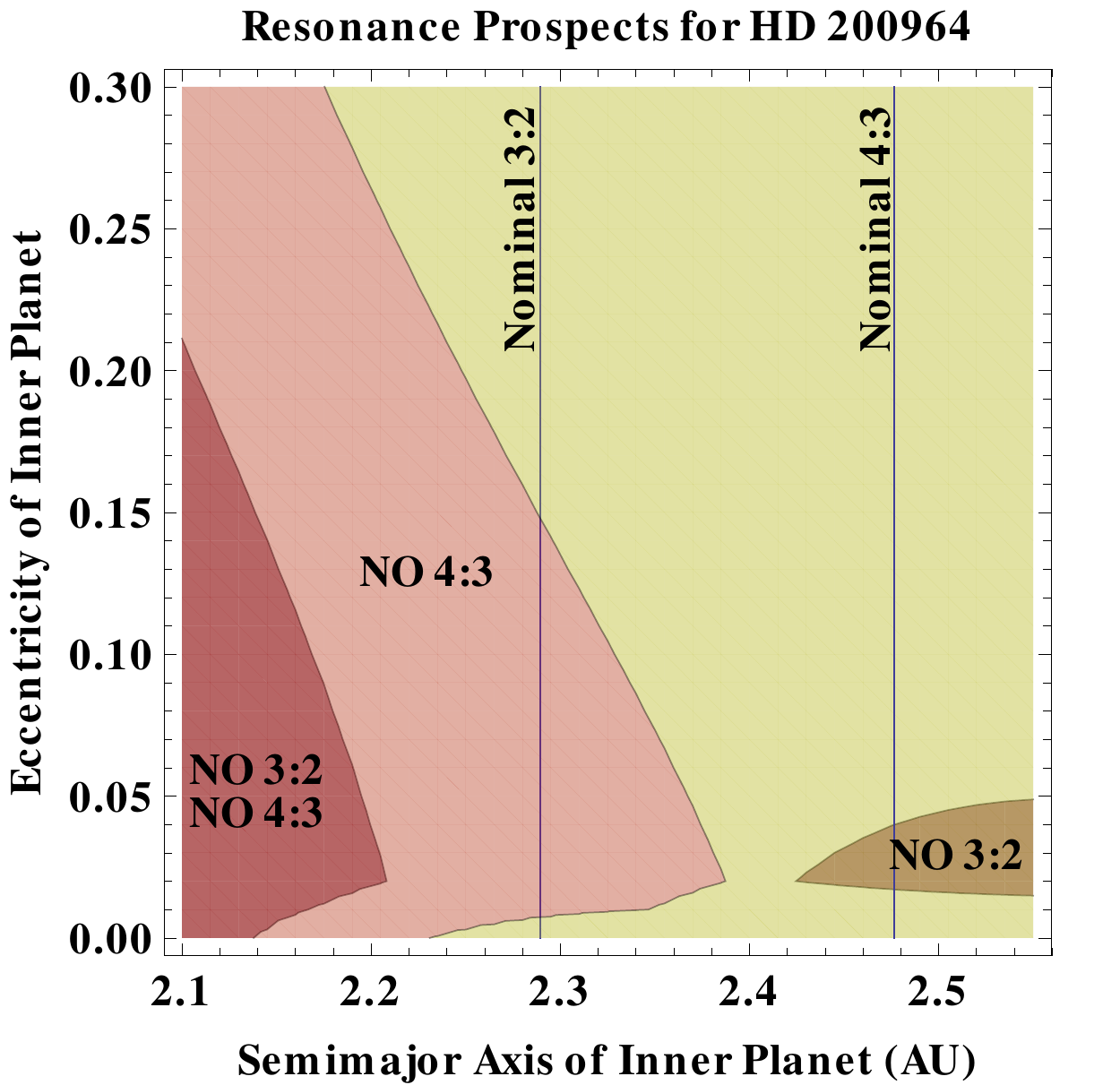,angle=0,width=0.3\textwidth}\label{fig:dis:a}}
\hspace{0.02\textwidth}
\subfigure[Same parameters as in Figure~\ref{fig:dis:a}.
The additional curves in the central panel are libration width estimates based on a single term in the disturbing function.  ]{
\psfig{figure=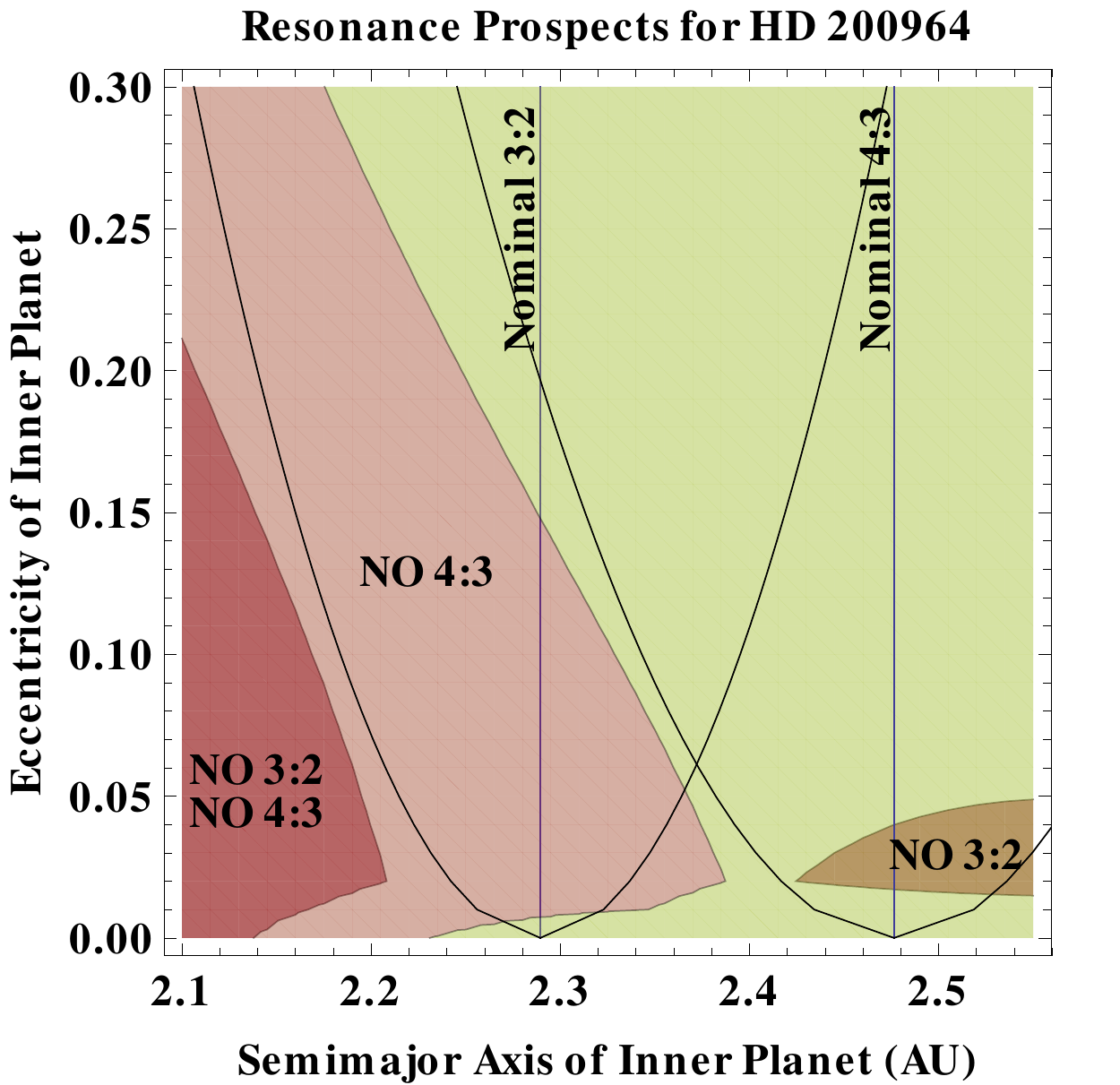,angle=0,width=0.3\textwidth}\label{fig:dis:b}}
\hspace{0.02\textwidth}
\subfigure[Inner planet is a test particle. All other parameters are the same as in Figure~\ref{fig:dis:a}.  ]{
\psfig{figure=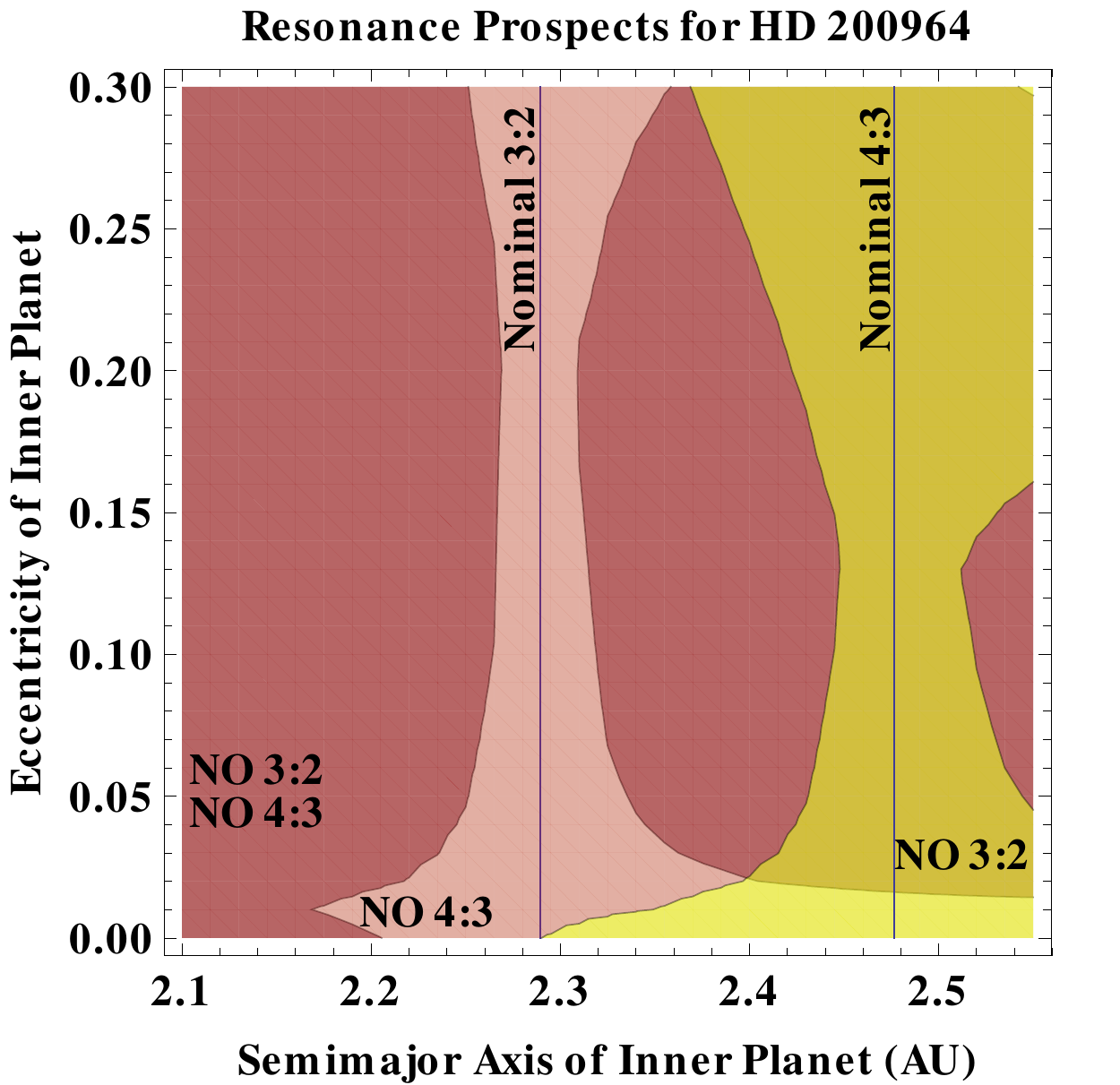,angle=0,width=0.3\textwidth}\label{fig:dis:c}}
\caption{Excluded resonant regions for the HD~200964 system.  
Plotted is the parameter space surrounding the $4$:$3$ and $3$:$2$ commensurability regions.
Regions labeled 'NO' indicate that a resonance cannot occur for the given commensurability.
}
\label{fig:stability:structure1}
\end{figure*}

Resonant theory has been applied to dynamical problems in the solar system with great success \citep[e.g.][]{solarsystemdynamics,morbidelli2002}.   
However, properties of extra-solar systems typically do not adhere to the approximations which can be adopted for the solar system \citep{beamic2003,verarm2007}.  
In particular, predicting the evolution of two massive bodies on non-circular orbits poses a rich dynamical challenge.  
Although much analytical progress has been made characterizing regimes of motion when these bodies are in the strong $2$:$1$ MMR \citep{beaetal2003,micetal2008a,micetal2008b}, few  investigators have modeled in detail other first-order resonances, partly because of their close proximity to the chaotic resonant overlap region \citep{wisdom1980,muswya2011}.
 
Further, detailed analyses of particular systems benefit from well-constrained observational data, showcasing another benefit of analyzing solar system bodies; the majority of confirmed extra-solar planet have unknown masses, bounded only from below.  
Orbital parameters for transiting systems are even less constrained, making detailed dynamical modeling almost impossible.
One approach to tackling these issues is to consider when a planetary system cannot be in resonance, by confirming that any potential librating angle must circulate \citep{verfor2011}.  
This procedure can be carried out analytically by using a disturbing function with a sufficient number of terms to accurately sample the desired system.
 
Here, we carry out the same procedure as in \cite{verfor2011} by using both the disturbing function\footnote{This is the same disturbing function which later appeared in \cite{solarsystemdynamics}.} from \cite{ellmur2000} to fourth-order in eccentricities and the analytical formulas from \cite{veras2007}.  
We assume coplanarity and use all resonant terms up to fourth-order (including the relevant 8:6,~12:9 and~16:12~terms) and secular terms up to fourth order.  
Additionally and separately we perform the same analysis for the 3:2~MMR, given its close proximity to the $4$:$3$ MMR.
 
The results are plotted in Figure~\ref{fig:stability:structure1}.
There are four different areas in these plots.
\begin{enumerate}
\item In the dark-orange region to the bottom-right labeled 'NO 3:2', no 3:2 solutions are permitted (while 4:3 solutions may or may not exist).
\item In the dark-red region to the far left, neither 3:2 nor 4:3 solutions are permitted. 
\item In the light-pink region labeled 'NO 4:3', no 4:3 solutions are permitted (while 3:2 solutions may or may not exist).
\item In the central light-orange region no definitive statement can be made to exclude either the 3:2 or 4:3 MMRs (but this does not equate to a statement that both can definitely exist).  
\end{enumerate}

Figure~\ref{fig:dis:a} illustrates the excluded 4:3~and 3:2~MMR regions for the HD~200964 system, assuming $m_{\star} = 1.44 M_{\odot}$, $m_1=1.8\,\Mjup$, $m_2=0.9\,\Mjup$ and fixed outer planet values of $a_2=3.0$~AU and $e_2=0.01$.  
This plot helps constrain the prospects for the system evolving in MMR given particular orbital parameters.  
Alternatively, if a MMR is assumed, then the plot helps constrain the planets' allowable orbital  parameters.  
Figure~\ref{fig:dis:b} over-plots a simple estimate of libration width from Eq.~4.46 of \citep{verarm2004} which assumes that just one disturbing function term is retained; this approach, often used in  the solar system, poorly reproduces the allowed resonant motions for this extra-solar system.  
Figure~\ref{fig:dis:c} demonstrates what the excluded regions look like in the limit of an inner planet mass of zero.  
In this case, both resonances become somewhat decoupled, and there is clear structure around  each nominal commensurability (see also Section~\ref{sec:stability:nbody}).  
This plot may be compared to Figure~8.7 of \cite{solarsystemdynamics}; differences arise because of the masses adopted here and the additional terms of the disturbing function used.

This set of plots illustrates the difficulty in restricting resonance phase space analytically for two massive exoplanets.
But because the system is only stable when it is protected by a resonance (see also next section) we can use such an analysis in the interpretation of orbital parameters which are weakly constrained from radial velocity data.

\subsection{Direct $N$-body simulations}\label{sec:stability:nbody}
\begin{figure}
\centering
\subfigure[\label{fig:widea2e2}Observed planetary masses in the~HD~200964 system: $m_1=1.99\,\Mjup$ and $m_2=0.90\,\Mjup$.]{
\psfig{figure=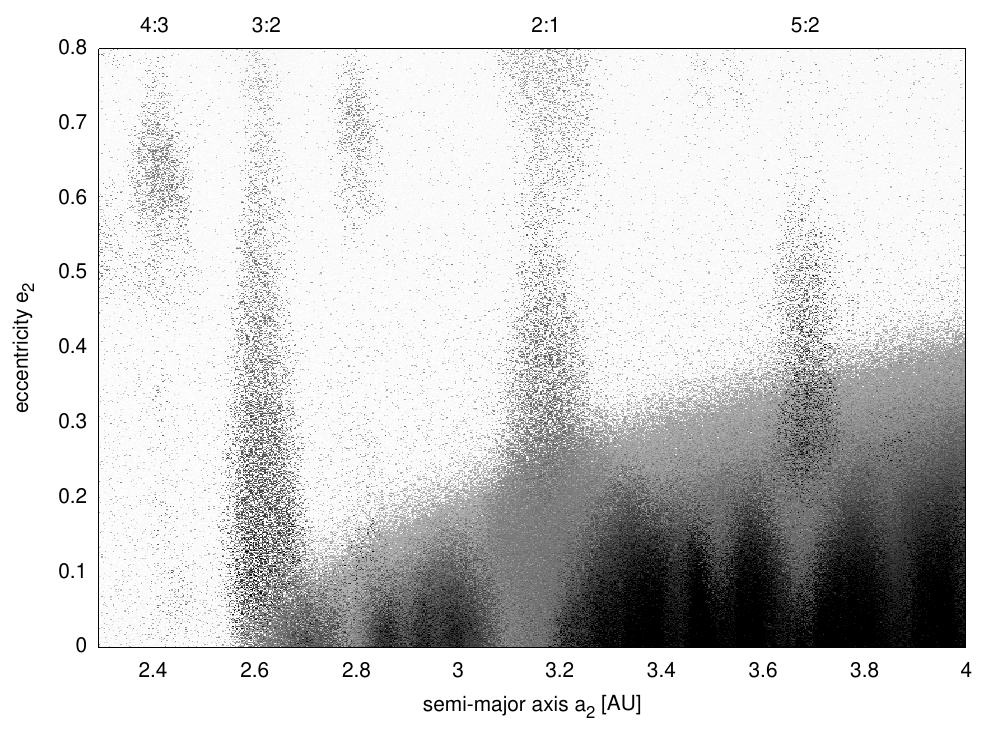,angle=0,width=1.0\columnwidth} }
\subfigure[\label{fig:widea2e23jup}\rev{Planetary masses larger then the observed minimum masses:} $m_1=m_2=3\,\Mjup$.]{
\psfig{figure=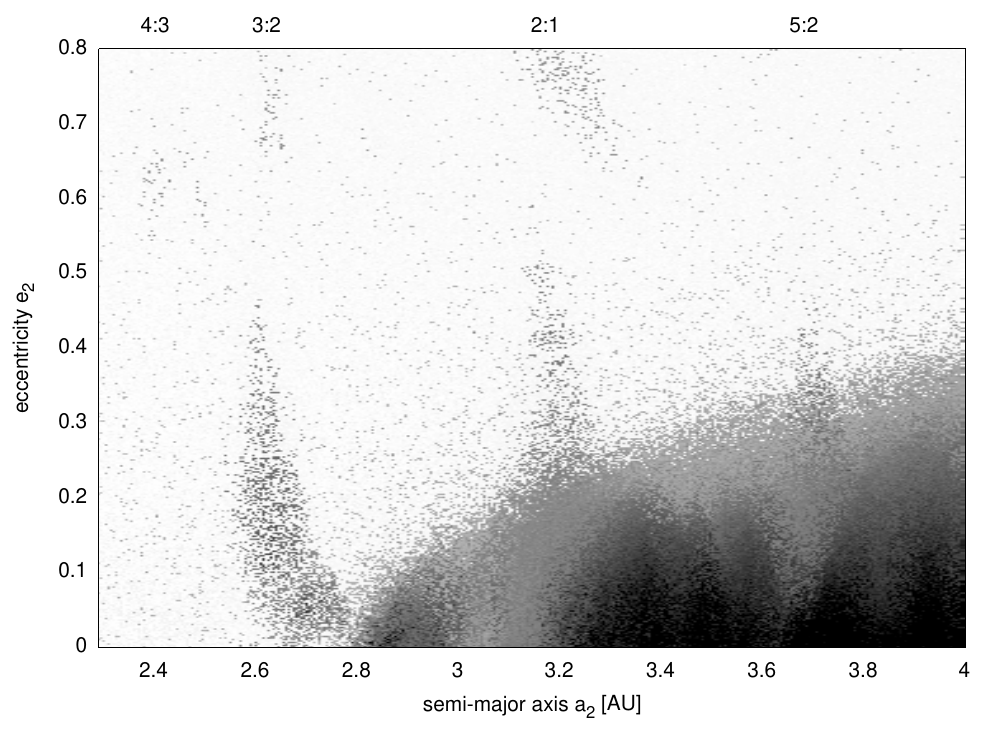,angle=0,width=1.0\columnwidth} }
\caption{Stability structure of the phase-space in a system with two planets.
The eccentricity and semi-major axis of the outer planet are varied.
The plotted quantity is the maximum Lyapunov exponent.
\rev{White regions are unstable, dark regions are stable.
The color-scale is the same for all stability plots.
Each pixel corresponds to a single $N$-body simulation.
}
}
\label{fig:stability:nbody:widea2e2}
\end{figure}

In the previous section we presented results of a resonant theory which was achieved by an expansion of the disturbing function. 
This allowed us to get an overview of the phase space structure of the HD~200964 system.
We also run direct $N$-body simulations of systems with two planets to investigate their stability.
The freely available code \texttt{REBOUND} \citep{ReinLiu2012} is used for all integrations in this section. 
We choose the Wisdom-Holman type integrator \citep{WisdomHolman1991} included in \texttt{REBOUND}.
\rev{To verify the results, we implemented a 15th order adaptive Radau integrator \citep{Everhart1985}.
We find that the results do not depend on the integrator or any numerical parameters such as the timestep.}

For each run, we add two shadow particles to the simulation in addition to the two planets.
This allows us to measure the long term stability of the system on short time-scales by calculating the maximum Lyapunov exponent.
To do that, the position and velocity of each shadow particle is initially set to those of the planets. They are then perturbed by a small amount. 
At regular intervals the rate of divergence between the shadow particle and the planet is measured and rescaled to the initial displacement.
By keeping track of the rescalings, we can estimate the maximum Lyapunov characteristic exponent. 
See \cite{Wisdom1983} for more details on the numerical algorithm.
\rev{If the numerically calculated Lyapunov timescale is smaller or comparable to the run time, we call the system system stable and unstable otherwise. 
The color scale in all stability plots was chosen to reflect this definition.
White regions are unstable, dark regions are stable and have a Lyapunov timescale that satisfies the above criteria.}

The Lyapunov exponent gives us a good overview of the parameter space. 
In Figure~\ref{fig:stability:nbody:widea2e2} we plot the results of $4\cdot10^5$ 
$N$-body simulations, each running for approximately $2\cdot10^4$~dynamical times. 
The mass of the central object is that of the star HD~200964. 
We show the results for three different planetary masses. 
For Figure~\ref{fig:widea2e2}, we use the observed masses of the HD~200964 system (see \Table{tab:rv}).
We repeat the same calculation with planets that each have 3 Jupiter masses in Figure~\ref{fig:widea2e23jup} and for Earth mass planets in Figure~\ref{fig:widea2e2earth}.
The initial semi-major axis and eccentricity of the inner orbit are $a_1=2, e_1=0.01$.
These values are the same for all simulations. 
The initial semi-major axis and eccentricity of the outer planet are varied and shown on the axis of the plot. 
All angles are chosen from a uniform distribution. 
The system is assumed to be coplanar\footnote{We discuss the formation of an inclined system in Appendix~\ref{app:nbody:inclined} but observe no qualitative difference to the coplanar case.}.

In Figure~\ref{fig:widea2e2}, one can see bands and islands of stability for systems in mean motion resonances. Examples are located at $a_2=2.62$~(3:2), $a_2=3.17$~(2:1) and $a_2=3.68$~(5:2). 
The location of the 4:3~mean motion resonance is also visible at $a_2=2.42$. 
This is a small stable island compared the other resonances mentioned. 
Also note that the eccentricity for these stable solutions is very high $e_2\sim0.64$.
However, this is also a function of the inner planet's eccentricity which has been kept fixed.
We present a slice of the parameter space in the $e_1$, $e_2$ plane in Appendix~\ref{app:widea1a2}.

\rev{The stable island that we attribute to the 4:3~MMR does not coincide with the best fit solution of \cite{JohnsonPayne2011}.
The eccentricities near the stable island are much higher than in the RV fit.
Nevertheless, we verified that their solution is indeed stable.
We conclude that their solution corresponds to fine tuned initial conditions.
 }

For systems with even higher mass planets than in the HD~200964 system, the stable regions get even smaller.
This can be verified in Figure~\ref{fig:widea2e23jup}.
Note that the reported masses for HD~200964 are the minimum masses and could be significantly larger if the system is inclined with respect to the line of sight.
However, this result suggests that the masses can't be much larger in order for the system to be stable.

Two additional plots showing a slice in the $a_1$, $a_2$ plane of the the parameter space are presented in Appendix~\ref{app:widea1a2}.

\section{Formation mechanisms}\label{sec:formation}
In this section we investigate potential methods for the formation of a pair of \rev{massive} planets in a 4:3 resonance. 
In Section~\ref{sec:formation:migration} we show that the long standing idea of convergent migration fails to produce closely packed resonances for massive planets.
We then consider the in-situ formation of planets in Section~\ref{sec:formation:insitu}, starting from small embryos in resonance which accrete mass from the protoplanetary disk.
In Section~\ref{sec:formation:scattering} scattering and simultaneous damping is considered as one possible alternative to the \textit{cold} formation scenarios mentioned above.
Finally, we discuss alternative formation scenarios in Section~\ref{sec:formation:alternatives}.

We acknowledge that other plausible mechanisms may exist, some of which are mentioned in the discussion section.

\subsection{Convergent migration in a disk}\label{sec:formation:migration}

Migration of planets through a disk depends on numerous parameters such as the planet and disk mass, disk viscosity, surface-density profile, disk scale height and the equation of state, to just mention a few. 
This allows for the possibility of convergent migration of planetary orbits, during which pairs of planets can pass through orbital period commensurabilities. 
If the convergent migration rate is sufficiently low, the planets can capture into resonance \citep{Gold65,ReinPapaloizouKley2010}.
In Section~\ref{sec:formation:migration:nbody} we will simplify the migration process by assuming that we can describe it with only one semi-major axis and one eccentricity damping time-scale per planet. 
This allows us to understand the physical processes at work during resonance capture, study a wide parameter regime and not get lost in the complicated details of planetary migration.
In Section~\ref{sec:formation:migration:hydro} we will then relax some of these simplifications and study the formation of a planetary system near a mean motion resonance using hydro-dynamical simulations.

\subsubsection{$N$-body simulations}\label{sec:formation:migration:nbody}
\begin{figure}
\psfig{figure=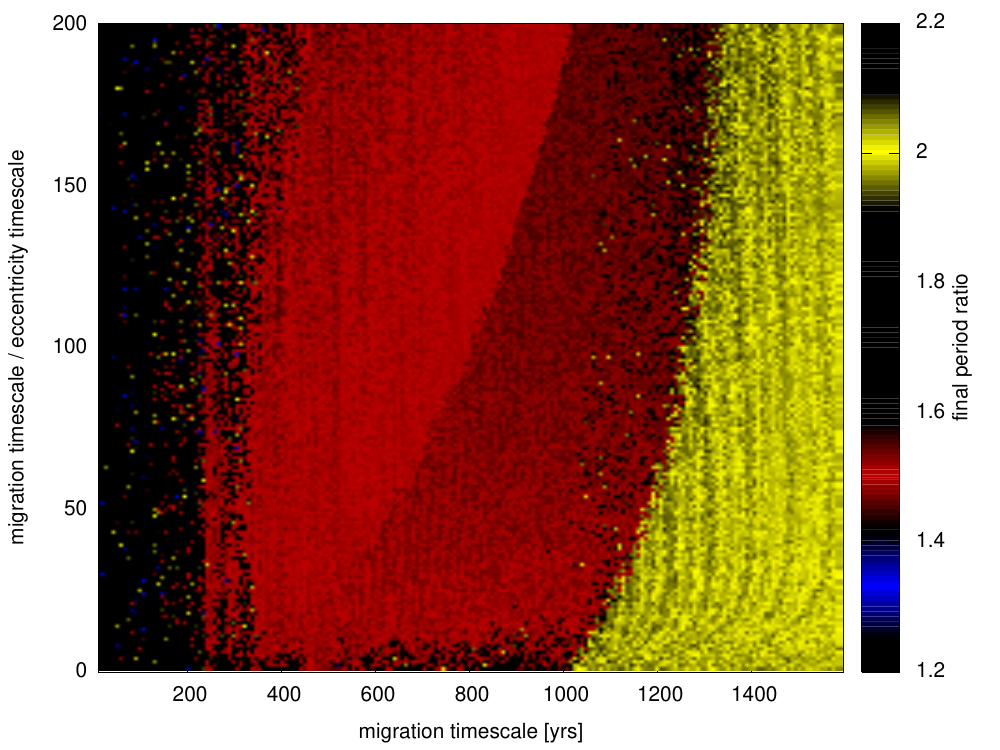,angle=0,width=1.0\columnwidth} 
\caption{ Final period ratios after convergent migration of two planets as a function of the migration and eccentricity damping timescales.  Observed planetary masses of the system HD~200964: $m_1=1.8 \Mjup$, $m_2=0.9 \Mjup$. \label{fig:formation:migration:nbody1} }
\end{figure}

We use the $N$-body code \texttt{REBOUND} \citep{ReinLiu2012} to model the orbital evolution of two massive planets.
We apply non-conservative forces to the outer planets which damp its semi-major axis and eccentricity \cite[see][for details on the implementation of the non-conservative forces]{LeePeale2002}. 
We also refer the reader to the work by \cite{ReinPapaloizouKley2010} on an investigation of the 3:2~resonance in the HD45364 system, which uses a similar methodology.
We experimented with applying non-conservative forces to the inner planet simultaneously, but did not see any qualitative difference and will not investigate this further.

Initially, the planets start far apart from each other on circular orbits.
The assumption of circular orbits is reasonable in the currently favored core accretion model of planet formation. 
Furthermore, the eccentricity damping time-scale is much shorter than the migration time-scale in a standard disk unless the eccentricity is very large \citep{Muto2011}. 
The planets are also assumed to be coplanar. 
Even though convergent migration and resonant capture can theoretically excite inclination \citep{TremaineYu2000}, the inclination damping time-scale is again much shorter than the migration time-scale as long as the planets are embedded in the disc, i.e. $i<10^\circ$ \citep{Rein2011}.
We also tried relaxing this condition and ran additional fully three dimensional simulations with finite relative initial inclination between the planets. 
The results are shown in Appendix~\ref{app:nbody:inclined}.
No significant changes can be observed. 
We therefore do not investigate this further. 

The outer planet migrates inwards on a time-scale $\tau_a$ and might get captured into a resonance with the inner planet.
In \Figure{fig:formation:migration:nbody1} we plot the final period ratio seen after the migration reached a quasi steady state and both planets migrate inwards self-similarly. 
The simulations are conducted with different semi-major axis damping rates and eccentricity damping rates which are the axes of the plot.
Each pixel represents one N-body simulation.
The color scheme has been chosen so that blue corresponds to a final period ratio close to 4:3, red is close to 3:2 and green is close to 2:1. 
The color black indicates that either planets are captured in another resonance or that at least one planet got ejected.
The latter is more common for short migration timescales,~$\tau_a$.

We find that it is essentially impossible to form systems in a 4:3 resonance via a simple convergent migration scenario with the observed mass ratio\footnote{Note that we use the minimum masses and ignore possibly even higher mass ratios if the system is inclined with respect to the line of sight.}. 
There are two fundamental problems.
First, a very high migration rate is required to allow the planets to pass through the 2:1 and 3:2~resonances.
Second, there is only a small stable island in parameter space around the 4:3 resonance and the migration scenarios tend not to put the planets into this specific location.

To understand the evolutionary track, recall the stability plot in \Figure{fig:widea2e2}. 
In the migration scenario the outer planet is initially on a circular orbit far away from the inner planet; that is the bottom right on the plot. 
Convergent migration brings the planets closer together.
On the plot, this corresponds to moving to the left along a horizontal line.
Depending on the migration speed, the planet feels the resonant interaction from the inner planet and starts gaining eccentricity, thus moving upwards in the plot. 
If the migration rate is slow enough, the planet will get captured into a resonance which will protect it from close encounters.
However, if the migration rate is so fast that the planet slips through all resonances, it ends up in an unstable regime (red region in the bottom left of the plot). 
In fact, it has to go through an unstable region to get to the stable island that corresponds to the 4:3~resonance. 
This is why such a smooth migration scenario cannot form tight resonances for massive planets.

\subsubsection{Hydro-dynamical simulations}\label{sec:formation:migration:hydro}
\begin{table*}
\begin{tabular}{lrrr|rr|rr|rr|rr}
\hline
Name &             $\Sigma_0$              & $\sigma$ & $f$ & $\nu$       & $h_0$ & $a_{1}$ & $a_{2}$ & $m_1$ & $m_2$ & $\dot{m}_1$ & $\dot{m}_2$\\
\hline
\hline
\texttt{vanilla}              & $ 2\cdot 10^{-4}$     & 0.5 &  0   & $10^{-5}$       & 0.04  & 1.8 & 2.4  & $1.22\cdot 10^{-3}$ & $5.97\cdot 10^{-4}$ &0&0    \\
\texttt{h0.07}                & $ 2\cdot 10^{-4}$     & 0.5 &  0   & $10^{-5}$       & 0.07  & 1.8 & 2.4  & $1.22\cdot 10^{-3}$ & $5.97\cdot 10^{-4}$ &0&0    \\
\texttt{sigma2}               & $ 4\cdot 10^{-4}$     & 0.5 &  0   & $10^{-5}$       & 0.04  & 1.8 & 2.4  & $1.22\cdot 10^{-3}$ & $5.97\cdot 10^{-4}$ &0&0    \\
\texttt{slope0}               & $ 2\cdot 10^{-4}$     & 0   &  0   & $10^{-5}$       & 0.04  & 1.8 & 2.4  & $1.22\cdot 10^{-3}$ & $5.97\cdot 10^{-4}$ &0&0    \\
\texttt{apart2}               & $ 2\cdot 10^{-4}$     & 0.5 &  0   & $10^{-5}$       & 0.04  & 1.8 & 2.9  & $1.22\cdot 10^{-3}$ & $5.97\cdot 10^{-4}$ &0&0    \\
\texttt{apart2\_sigma8}       & $16\cdot 10^{-4}$     & 0.5 &  0   & $10^{-5}$       & 0.04  & 1.8 & 2.9  & $1.22\cdot 10^{-3}$ & $5.97\cdot 10^{-4}$ &0&0    \\
\texttt{outer0.5}             & $ 2\cdot 10^{-4}$     & 0.5 &  0   & $10^{-5}$       & 0.04  & 1.8 & 2.4  & $1.22\cdot 10^{-3}$ & $3.00\cdot 10^{-4}$ &0&0    \\
\texttt{seed\_accretion}      & $ 2\cdot 10^{-4}$     & 0.5 &  0   & $10^{-5}$       & 0.04  & 1.8 & 2.3  & $1.22\cdot 10^{-3}$ & $1.00\cdot 10^{-5}$ &0&1    \\
\texttt{seed\_close}          & $ 2\cdot 10^{-4}$     & 0.5 &  0   & $10^{-5}$       & 0.04  & 1.8 & 2.05 & $1.22\cdot 10^{-3}$ & $3.00\cdot 10^{-5}$ &0&1    \\
\texttt{alpha4}               & $ 2\cdot 10^{-4}$     & 0.5 &  0   & $4\cdot10^{-5}$ & 0.04  & 1.8 & 2.4  & $1.22\cdot 10^{-3}$ & $5.97\cdot 10^{-4}$ &0&0    \\
\texttt{close}                & $ 2\cdot 10^{-4}$     & 0.5 &  0   & $10^{-5}$       & 0.04  & 1.8 & 2.1  & $1.22\cdot 10^{-3}$ & $5.97\cdot 10^{-4}$ &0&0    \\ 
\texttt{seed\_accretion\_both}             & $ 2\cdot 10^{-4}$     & 0.5 &  0   & $10^{-5}$       & 0.04  & 1.8 & 2.3  & $      1.00\cdot10^{-5}$ & $1.00\cdot 10^{-5}$ &1&1    \\
\texttt{seed\_accretion\_both\_moremassive}     & $ 2\cdot 10^{-4}$     & 0.5 &  0   & $10^{-5}$       & 0.04  & 1.8 & 2.3  & $  1.00\cdot        10^{-4}$ & $1.00\cdot 10^{-4}$ &1&1    \\
\texttt{seed\_accretion\_apart}             & $ 2\cdot 10^{-4}$     & 0.5 &  0   & $10^{-5}$       & 0.04  & 1.8 & 3.0  & $1.22\cdot 10^{-3}$ & $1.00\cdot 10^{-5}$ &0&0    \\
\texttt{sigma8}               & $16\cdot 10^{-4}$     & 0.5 &  0   & $10^{-5}$       & 0.04  & 1.8 & 2.4  & $1.22\cdot 10^{-3}$ & $5.97\cdot 10^{-4}$ &0&0    \\
\texttt{sigma8\_outer0.1}            & $16\cdot 10^{-4}$     & 0.5 &  0   & $10^{-5}$       & 0.04  & 1.8 & 3.0  & $1.22\cdot 10^{-3}$ & $5.97\cdot 10^{-5}$ &0&0    \\
\texttt{sigma4\_outer0.1}             & $ 8\cdot 10^{-4}$     & 0.5 &  0   & $10^{-5}$       & 0.04  & 1.8 & 3.0  & $1.22\cdot 10^{-3}$ & $5.97\cdot 10^{-5}$ &0&0    \\
\texttt{sigma2\_outer0.1}             & $ 4\cdot 10^{-4}$     & 0.5 &  0   & $10^{-5}$       & 0.04  & 1.8 & 3.0  & $1.22\cdot 10^{-3}$ & $5.97\cdot 10^{-5}$ &0&0    \\
\texttt{sigma4\_outer0.1\_alpha4}           & $ 8\cdot 10^{-4}$     & 0.5 &  0   & $4\cdot10^{-5}$ & 0.04  & 1.8 & 3.0  & $1.22\cdot 10^{-3}$ & $5.97\cdot 10^{-5}$ &0&0    \\
\texttt{sigma4\_outer0.1\_h0.03}           & $ 8\cdot 10^{-4}$     & 0.5 &  0   & $10^{-5}$       & 0.03  & 1.8 & 3.0  & $1.22\cdot 10^{-3}$ & $5.97\cdot 10^{-5}$ &0&0    \\
\texttt{sigma2\_m0.1}               & $ 4\cdot 10^{-4}$     & 0.5 &  0   & $10^{-5}$       & 0.04  & 1.8 & 3.0  & $1.22\cdot 10^{-4}$ & $5.97\cdot 10^{-5}$ &0&0    \\
\texttt{sigma4\_m0.1}               & $ 8\cdot 10^{-4}$     & 0.5 &  0   & $10^{-5}$       & 0.04  & 1.8 & 3.0  & $1.22\cdot 10^{-4}$ & $5.97\cdot 10^{-5}$ &0&0    \\
\texttt{sigma4\_outer0.1\_flare}           & $ 8\cdot 10^{-4}$     & 0.5 & -0.5 & $10^{-5}$       & 0.04  & 1.8 & 3.0  & $1.22\cdot 10^{-3}$ & $5.97\cdot 10^{-5}$ &0&0    \\
\texttt{sigma4\_m0.1\_flare}             & $ 8\cdot 10^{-4}$     & 0.5 & -0.5 & $10^{-5}$       & 0.04  & 1.8 & 3.0  & $1.22\cdot 10^{-4}$ & $5.97\cdot 10^{-5}$ &0&0    \\
\texttt{seed\_inresonance}     & $ 2\cdot 10^{-4}$     & 0.5 &  0   & $10^{-5}$       & 0.04  & 2.0 & 2.427  & $3.00\cdot 10^{-5}$ & $3.00\cdot 10^{-5}$ &1&1    \\
\texttt{seed\_inresonance\_slow}     & $ 2\cdot 10^{-4}$     & 0.5 &  0   & $10^{-5}$       & 0.04  & 2.0 & 2.427  & $3.00\cdot 10^{-5}$ & $3.00\cdot 10^{-5}$ &0.1&0.1    \\
\texttt{seed\_inresonance\_outer4}     & $ 2\cdot 10^{-4}$     & 0.5 &  0   & $10^{-5}$       & 0.04  & 2.0 & 2.427  & $3.00\cdot 10^{-5}$ & $1.20\cdot 10^{-4}$ &1&1    \\
\texttt{seed\_inresonance\_slope0}     & $ 2\cdot 10^{-4}$     & 0 &  0   & $10^{-5}$       & 0.04  & 2.0 & 2.427  & $3.00\cdot 10^{-5}$ & $3.00\cdot 10^{-5}$ &1&1    \\
\texttt{seed\_inresonance\_slope0\_veryslow}     & $ 2\cdot 10^{-4}$     & 0 &  0   & $10^{-5}$       & 0.04  & 2.0 & 2.427  & $3.00\cdot 10^{-5}$ & $3.00\cdot 10^{-5}$ &0.01&0.01    \\
\hline
\end{tabular}
\caption{\label{tab:hydrosimulations}Parameters used in the hydrodynamic simulations. 
All parameters are given at $1\mathrm{AU}$, they may be different at other locations due to gradients in the disk.
The first column gives the name of the simulation. 
The second column lists the surface density in units of $\Msun/\mathrm{AU}^2$.
The third and fourth columns give the density gradient ($\Sigma = \Sigma_0\,r^\sigma$) and the flaring index of the disk ($h=h_0\,r^f$).
The fifth column gives the kinematic viscosity in units of $2\pi \mathrm{AU}^2/\mathrm{yr}$.
The sixth column gives the aspect ratio of the disk.
The seventh and eighth columns list the initial semi-major axis of the inner and outer planet respectively.
The ninth and tenth columns give the initial mass of the inner and outer planet respectively. 
The eleventh and twelfth columns indicate if the planets are allowed to accrete mass (using the Kley formalism).
}
\end{table*}

\begin{figure*}
\centering
\psfig{figure=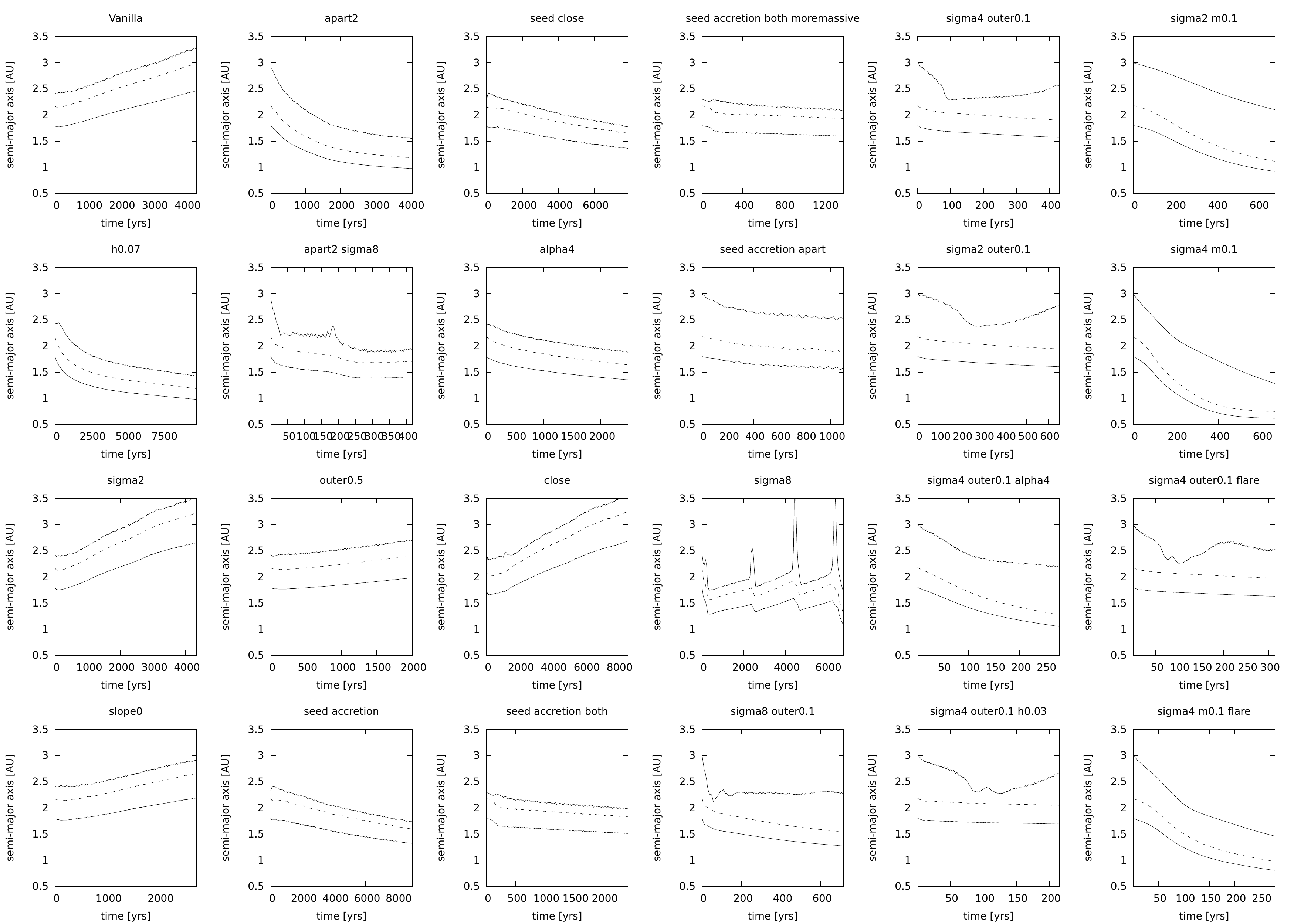,angle=0,width=\textwidth}
\caption{
\label{fig:formation:migration:hydro1}
Results of hydro-dynamical simulations. 
The solid lines show the semi-major axes of the inner and outer planet. 
The dashed line shows the semi-major axis corresponding to an outer 4:3 MMR with the inner planet.
}
\end{figure*}

\begin{figure*}
\centering
\psfig{figure=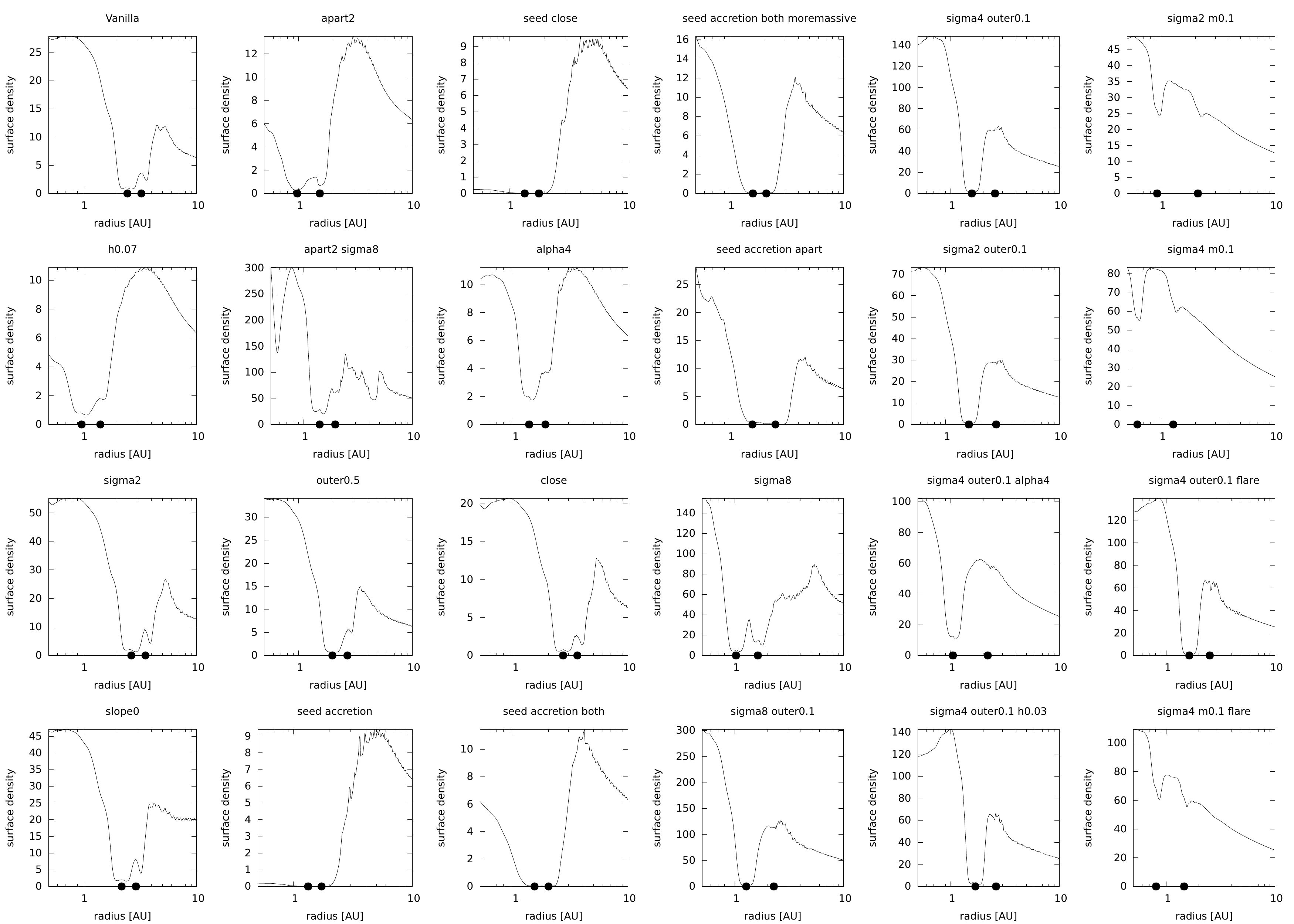,angle=0,width=\textwidth}
\caption{
\label{fig:formation:migration:hydro2}
Results of hydro-dynamical simulations. 
The plots show the gas surface density and the position of the two planets at the end of the simulation.
Note that the final time varies significantly between different simulations.
}
\end{figure*}

We now go beyond the simple $N$-body model and use the two dimensional hydrodynamic code \texttt{FARGO} \citep{Masset00} to perform simulations of two gravitationally interacting planets that also undergo interactions with an accretion disk.
This setup is used to further test the rapid migration hypothesis investigated with $N$-body simulations and parameterized migration forces in Section~\ref{sec:formation:migration:nbody}.
Again, the approach is conceptually similar to that of \cite{ReinPapaloizouKley2010}.
We are in particular looking for more complicated effects such as varying migration rates that might be overlooked by running simplified $N$-body simulations. 

Our simulations use a cylindrical grid with $N_\phi=628$ and $N_r=512$. 
The radial extent of the computational domain goes from $0.5\,\mathrm{AU}$ to $5\,\mathrm{AU}$. 
We use non-reflecting boundary conditions to minimize the effects from the finite domain size.
Both the disk and planet properties are varied in the simulations and are listed in \Table{tab:hydrosimulations}.
We run a total of over 50 different simulations\footnote{Not all are presented in the table.}.
All planets except those in simulations labeled \texttt{inresonance} (see Section~\ref{sec:formation:insitu} for a discussion of those) start initially on circular orbits.
Some of the simulations start with small mass planets and allow for accretion of mass from the proto-stellar disk.

None of the simulation listed in \Table{tab:hydrosimulations} results in a stable 4:3~resonance. 
The simulations either capture into a 2:1~or 3:2~resonance or scatter due to close encounters. 
The result is therefore in perfect agreement with the $N$-body simulations of Section~\ref{sec:formation:migration:nbody}.

In Figures \ref{fig:formation:migration:hydro1} we show the evolution of the semi-major axis of both planets as solid lines.
The dashes line shows the nominal position of the 4:3 resonance of the inner planet and is thus the semi-major axis we are so desperately trying to reach.
As can be seen easily, we do not achieve this. 
In Figures \ref{fig:formation:migration:hydro2} we also plot the surface density profile of the disk at the end of the simulation.
Note that the final simulation time varies significantly between the simulations. 
Whenever a steady or adiabatic state was achieved, we decided to stop the simulation.
There are several physical reasons why we cannot get the planets in a 4:3 resonance with these kind of migration scenarios.

First, the planets are very massive and therefore the resonances are strong. 
This prefers more widely separated resonances as has already been shown by the $N$-body simulations above.
It can be verified in the simulations labeled \texttt{vanilla}, \texttt{h0.07}, \texttt{slope0} and many others. 

\rev{Second, if we force the planets to have an extremely rapid (and most likely completely unphysical) migration rate, the planets do indeed get closer initially.
This can be seen in the simulation \texttt{sigma8} that has a very massive disk which leads to a fast migration rate. 
Once in resonance, the planets start migrating outwards again.
The same happens in the Grand Tack Scenario \citep{Walsh2011}. 
After a while the outer planets starts migrating out very quickly due to the heavy disk and a Type III migration regime.
This breaks the resonance.
Shortly after the resonance is lost, the planets start moving in again, recapture in resonance and the whole cycle repeats.
We never observe the planets capturing in a 4:3 resonance in this process.
}

Third, the inner planet is massive and will open a gap in any reasonable disk model. 
The location of the 4:3 resonance is a factor of 1.2 away from the inner planet in terms of its semi-major axis. 
Note that the gap cannot be smaller than a few scale heights or Hill radii \citep{Crida2006}. 
\rev{However, depending on the precise planet mass and disk model the outer planet might not open a gap on its own. 
In that case it might get trapped at the gap edge.
Migration stops or at least slows down, preventing the outer planet getting closer.
}

Fourth, as \cite{Podlewska2011} point out, the interaction of the outer planet with the waves launched by the inner planet can cause additional torques \rev{if the outer planet does not open a gap on its own}. 
This surfing effect tends to move the outer planet further away. 
This can be seen in simulations labeled \texttt{sigma4\_outer0.1\_h0.03}, \texttt{sigma8\_outer0.1}, \texttt{sigma4\_ouer0.1} and a few others.

Of course this cannot be a complete survey of all possible scenarios even though we ran over 50 hydro-dynamical simulations.
Also, there are significant limitations and errors associated with such a numerical scheme. 
The resolution has been kept fixed and we did not check for convergence in every simulation. 
However, it seems very unlikely that this would lead to a completely different picture as not even one of our simulations comes even close to capturing the two planets in the 4:3~resonance for a significant amount of time.

To conclude, the results of our hydro-dynamical simulations are in agreement with the $N$-body simulations. 
They do not allow the capture of planets in the 4:3 resonance.
In fact, due to gap opening and the surfing effect the problem of forming the resonance seems even harder to overcome than we would have estimated from $N$-body simulations.

\subsection{In-situ formation}\label{sec:formation:insitu}
\begin{figure*}
\centering
\psfig{figure=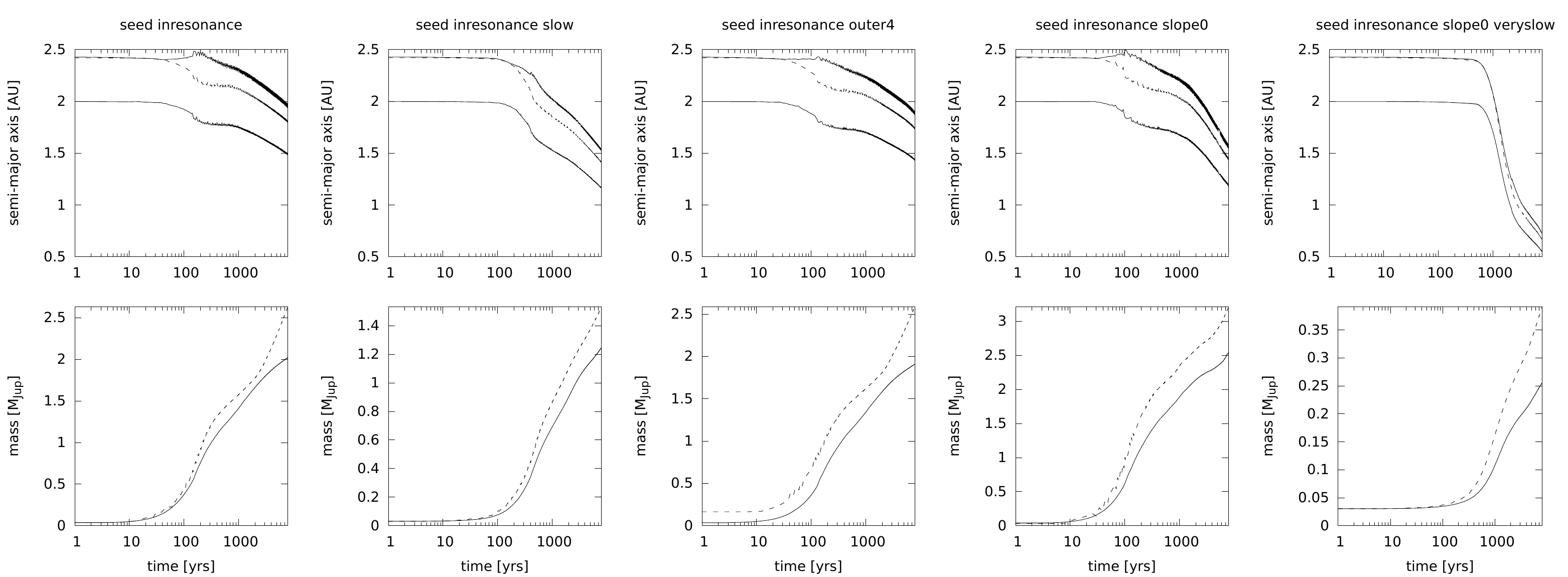,angle=0,width=\textwidth}
\caption{
\label{fig:formation:insitu}
Results of hydro-dynamical simulations of in-situ formation. 
\textit{Top row:} The solid lines show the semi-major axes of the inner and outer planet. 
The dashed (middle) line shows the semi-major axis corresponding to an outer 4:3 MMR with the inner planet.
\textit{Bottom row:} The mass of the planets as a function of time. The dashed line corresponds to the outer planet.
}
\end{figure*}
While the presence of a MMR is generally taken as being the signature of convergent migration (see Section~\ref{sec:formation:migration}), perhaps the most conceptually simple scenario to consider is that the two planets in the 4:3 resonance formed directly from embryos with 4:3 period ratios. 
At early times with much lower masses, their mutual interactions would have been much weaker and some period of relatively unperturbed growth could have taken place.
There is then the possibility that the pair of 4:3 planets which formed in-situ could have migrated together through the disk to the currently-observed locations. 

We test this idea in a similar framework to that used in the last section. 
However, we do not start planets far apart from each other, but start them directly in the 4:3~MMR.
To do this, we run an $N$-body simulation using \texttt{REBOUND} with two ten Earth mass embryos.
We choose a migration time-scale which results in a capture in the 4:3~MMR (see Section~\ref{sec:formation:migration:nbody}).
After the planets have reached an equilibrium state and migrate together adiabatically, we switch to a full hydrodynamic simulation as in Section~\ref{sec:formation:migration:hydro}.
We let the embryos interact with the disk and accreted mass.
\rev{The accretion not only changes the mass but adds an additional component to the torque that is felt by the planet.}
The interaction with the disk is turned on adiabatically over several orbits.
We vary the accretion rate, the mass of the outer planet, and the slope of the initial surface density profile.
The full list of initial conditions for these simulations is shown at the bottom of \Table{tab:hydrosimulations} under \texttt{seed\_inresonance}, \texttt{seed\_inresonance\_slow}, \texttt{seed\_inresonance\_slope0}, \texttt{seed\_inresonance\_outer4} and \texttt{seed\_inresonance\_slope0\_veryslow}.

We plot the evolution of the semi-major axis and the mass of both planets in these simulations in \Figure{fig:formation:insitu}. 
One can see that the 4:3~MMR is lost a few hundred to a few thousand years after the planets are inserted into the disk.
The path to loosing the resonance is as follows.
While the planets are accreting mass, they are also migrating. 
They migrate initially in type-I, later in type-II when their mass is sufficiently large.
\rev{As the planets grow in mass, their dynamical interaction becomes stronger. 
The eccentricities of both planet rise. 
Eventually the resonance is lost.}
None of the planets gets ejected, but the interactions push them several Hill radii apart.
Note that slowing down the process, as in simulation \texttt{seed\_inresonance\_slope0\_versyslow} does not prevent the resonance breaking.
It merely delays it. 

There are clearly many more parameters that one could vary. 
However, it seems unlikely that we can keep the planets locked in resonance while they grow in mass by more than one order of magnitude.
This is because the mass that the embryos accrete has to come from the proto-planetary disk.
The planets always interact with the disk, exchange angular momentum and start migrating.

\subsection{Scattering and simultaneous damping}\label{sec:formation:scattering}
It has been shown in previous studies of planet-planet scattering \citep{Chatterjee2008,Raymond2008} that the formation of mean motion resonances in the aftermath of scattering between planets is possible, but that this is a relatively rare event (less than 1\% level). 
The low probability of capture into resonance in these investigations is likely a simple consequence of the fact that these simulations are conducted in the absence of any disk-gas damping (or other dissipative forces) and hence there is no means for the system to damp into resonance. 

When planet-planet scattering simulations are conducted in the presence of a disk gas \citep[e.g. ][]{Matsumura2010b,Moeckel2012a}, the dissipative gas component means that a much higher fraction of systems are observed to form MMRs of a wide variety of orders (see their Table 4). 
However, \citet{Matsumura2010b} did not explicitly observe the creation of 4:3 resonances.
We therefore investigate the scattering scenario further to understand whether such systems are simply rare (and thus statistically unlikely to be seen in their simulations), or whether the formation of a 4:3 resonance in this manner is essentially impossible.  

We start from the assumption that an inner planet exists in an unperturbed, approximately circular orbit at a semi-major axis $a_1 \sim 1.6\,\mathrm{AU}$. 
External to this, we assume that there existed an outer ($a \gg 2\,\mathrm{AU}$) population ($N \geq 2$) of massive planets which became unstable and scattered a Jupiter-mass planet onto an orbit with a pericenter at $q_2\sim 1.9\,\mathrm{AU}$ (the radial location of the 4:3 resonance with the inner planet at $1.6\,\mathrm{AU}$). 
We do not perform the lengthy, chaotic scattering simulations of this outer population, but rather assume that a scattering event has occurred with the outcome being that a $0.9\,\Mjup$ planet has been placed onto an orbit with $q_2 \sim 1.9\,\mathrm{AU}$. 

We consider that the outer body interacts with a remnant gas disk of some form, causing its orbit to damp, reducing in eccentricity and semi-major axis. 
In some fraction of systems with suitable damping terms, while the outer body is simultaneously circularizing and migrating inwards, the two planets may become trapped into a 4:3~MMR.

To investigate this scenario we perform $N$-body integrations which include the effect of gas damping to investigate whether the system described above can be captured into a 4:3~resonance.
The full details of the manner in which the simulations are specified are detailed in the following section.

\subsubsection{Simulation methodology}\label{sec:formation:scattering:method}
We initialize each simulation with conditions approximating some precursor of the HD~200964 system. 
The central star has a mass of $M_{\star}\sim 1.44M_{\odot}$.
The inner planet has a of $m_1 = 1.8 \Mjup$ and is located at $a_1 = 1.6\,\mathrm{AU}$ on a circular orbit, $e_1 = 0$.
The outer planet has a mass of $m_2 = 0.9 \Mjup$.
The pericenter $q_2$ is randomly chosen from a flat distribution $1.6 < q_2 < 2.2\,\mathrm{AU}$.
Note that the pericenter lies close to the location of the 4:3 resonance with the inner planet.
The outer planet's eccentricity is also randomly drawn from a flat distribution $e_{crit} < e_2 < 0.9$, where $e_{crit} = 1 - q_2/2.54$.
Thus the outer planet is constrained to start with a semi-major axis outside the location of the 2:1 resonance with the inner planet (at $2.54\,\mathrm{AU}$).

We parameterize the damping in the same manner as in Section~\ref{sec:formation:migration:nbody}, i.e. following the approach of \cite{LeePeale2002} by calculating the Jacobi orbital elements.
The semi-major axis $a$ and the eccentricity $e$ are then directly damped in the parameterized form $\frac{\dot{e}}{e}=K\frac{\dot{a}}{a}$, where the damping rate $\frac{\dot{a}}{a}$ and the ratio of eccentricity damping to semi-major axis damping, $K$, are varied as free parameters. 
We implement this damping in a modified version of the the Mercury integration package \citep{Chambers1999}.
Each simulation uses the hybrid-symplectic integrator in Mercury, and integrates the systems for a total time period $T_\mathrm{final}=100 a/\dot{a}$.

We perform 100 versions for each of the following 64 parameter combinations, thus performing a total of 6400~individual simulations. 
The damping rates $\dot{a}/a$ that we choose are $10^{-6}~\mathrm{yr}^{-1}$, $10^{-5}~\mathrm{yr}^{-1},$ $10^{-4}~\mathrm{yr}^{-1}$ and $10^{-3}~\mathrm{yr}^{-1}$.
We investigate four values of $K$, $0.01,0.1,1$ and $10$.
We perform simulations in which the damping occurs on either only the outer planet or both planets.
We also vary the location within which the gas damping operates: either over a wide range $r < 50\,\mathrm{AU}$, or in a small disk $r < 2.5\,\mathrm{AU}$. 
The latter implies that the outer planet will initially only be damped when very close to pericenter.

\subsubsection{Simulation results}\label{sec:formation:scattering:results}
\begin{figure}
\centering
\begin{tabular}{c}
\includegraphics[width=0.9\columnwidth,trim= 2.5cm 2.5cm 10.5cm 2.5cm]{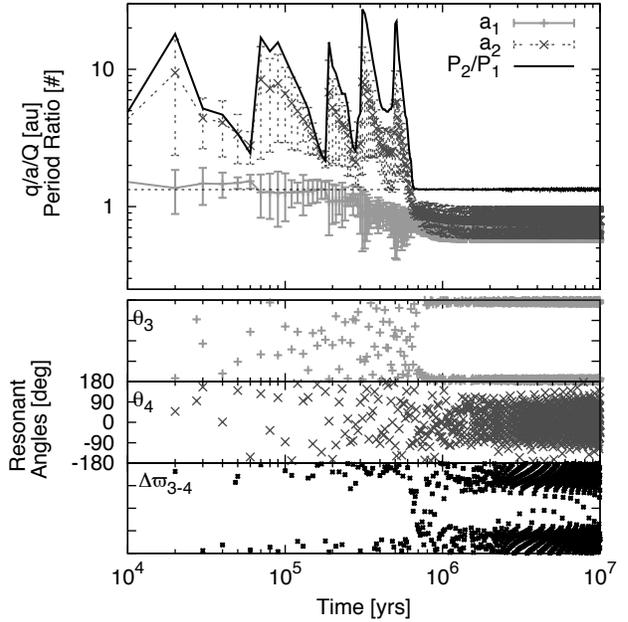}\\
\end{tabular}
\caption{Evolution of two planets in the aftermath of a scattering event with a third planet (not modeled). The top plot shows the semi-major axes, the pericenters and apocenters as well as the period ratio as a function of time. The bottom plot shows the 4:3 resonance angles.
}
\label{fig:formation:scattering:results1}
\end{figure}

\begin{figure}
\centering
\begin{tabular}{c}
\psfig{figure=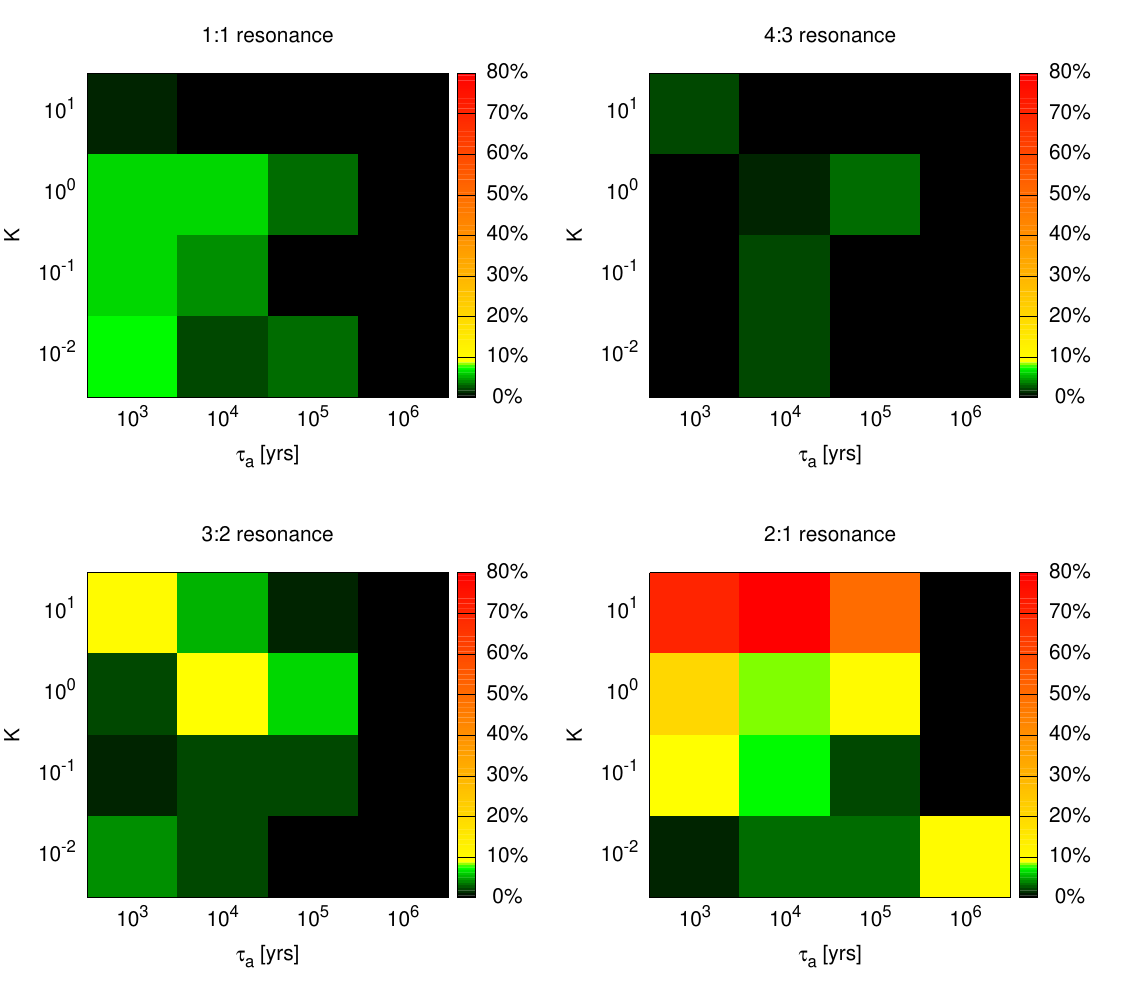,width=1.0\columnwidth}\\
\end{tabular}
\caption{The fraction of scattering simulations which were subsequently captured into various mean motion resonances as a function of the damping parameters $K$ and $\tau_a$.
}
\label{fig:formation:scattering:results2}
\end{figure}

We plot an example of a successful damping simulation in which an initially highly eccentric planet is subsequently made to damp and migrate into a 4:3~MMR with the inner planet in \Figure{fig:formation:scattering:results1}.
In the example shown, the outer planet initially had orbital parameters such that $a_2 = 3.59$, $e_2 = 0.53$ and therefore $q_2 = 1.66$.
The damping model was such that $\dot{a}/a = 10^{-5}\;\mathrm{yr}^{-1}$ and $K=1$, with this damping operating on the outer body only, for all $r < 50\,\mathrm{AU}$.
Over the first $t\sim 6\times 10^5\,\mathrm{yr}$ of the simulation the outer planet can be seen to continuously suffer close encounters with the inner planet (kicking both semi-major axes and eccentricities to new values), while also experiencing rapid damping and inward migration due to the gas interactions. 
Finally at $t\sim 6\times 10^5\,\mathrm{yr}$ the damping component wins and the planet definitively captures into a 4:3 resonance with the inner planet.
We plot the resonant angles $\theta_3,\,\theta_4$ and $\theta_{3-4}$ in bottom of \Figure{fig:formation:scattering:results1}.
They are defined as
\begin{eqnarray}
\theta_3 &=& 4(\lambda_2 - \varpi_2) - 3(\lambda_1 - \varpi_1) + 3(\varpi_2-\varpi_1)        \nonumber \\
\theta_4 &=& 4(\lambda_2 - \varpi_2) - 3(\lambda_1 - \varpi_1) + 4(\varpi_2-\varpi_1) \label{EQN:RES:ANGLES} \\
\theta_{3-4} &=& \theta_3 - \theta_4,        \nonumber
\end{eqnarray}
where $\lambda_i$ and $\varpi_i$ are the mean longitude and longitude of periapse of the $i^{\mathrm{th}}$ planet, respectively.
We observe that the resonant angle $\theta_3$ is librating.

The results of all 6400 individual simulations performed are shown in \Figure{fig:formation:scattering:results2}.
The plots show the fraction of systems in a certain resonance as a function of the damping timescales.
We marginalize over the other parameters.
Approximately $1\%$ of all systems are captured into the 4:3 resonance.
Shorter damping time-scales in both semi-major axis and eccentricity tend to result in higher capture fractions into the 4:3 resonance.

However, we note that if this model is at all realistic, then we also predict that the fraction of systems occupying a 1:1 resonance would be higher than the fraction of systems in a 4:3 resonance.
One can see from \Figure{fig:formation:scattering:results2} that the fraction of systems in a 1:1 resonance is more than 3 times as large as the fraction of systems in a 4:3 resonance.
An example of a capture into a 1:1 resonance is illustrated in Appendix \ref{app:onetoone}. 
Such resonances were examined by \citet{LC2002} and shown to be detectable by Kepler \citep{FG2006}. 
This is consistent with simulations of multiple small mass planets embedded in a gas disk performed by \cite{CresswellNelson2006}.
These authors find that 20\% of their runs produce co-orbital planets which survive until the end of their integration.

We reiterate the point that these simulations were deliberately initialized into a finely-tuned configuration that we hoped would lead to capture into a 4:3 resonance. 
Moreover, the model we use for our gas damping is a simple, parameterized one.
We thus cannot claim to have fully explored whether such a scenario would be common (or even at all possible).
Further study should address the likelihood of the initial scattering occurring.
This is a notoriously hard task because the initial mass function of planets (and their position) is not well understood.
Also, the realism of the damping scenario can only be checked when full hydro-dynamical simulations are employed. 
However, it is currently unfeasible to run thousands of hydrodynamic simulations for such a long time span ($t\sim 10^6$ years).

But what we can say is that the 4:3 resonance is clearly not a strong attractor in scattering simulations. 
Our results predict a large number of 1:1 resonances which are not observed. 
The overall likelihood to form a 4:3 resonance in our fine tuned simulations is only 1\%.
In a less fine tuned setup this rate would be much less then 1\% and therefore much less than the observed occurrence rate.

\subsection{Alternative means of forming planets in a 4:3~resonance}\label{sec:formation:alternatives}
One possible means of generating two planets in a 4:3 MMR is through the breaking of a resonant chain initially involving at least three planets.  
Such a resonant chain has been observed in planet-disk simulations of small mass planets by \cite{CresswellNelson2009}.
The Kepler sample provides an interesting testing-ground for these systems \citep{Ford2012}.

There are few investigations dedicated to measuring four-body resonances outside of the well-studied Laplace resonance involving Io, Europa and Ganymede.  
\citet{1998CeMDA..71..243N,1998AJ....116.3029N} do so, but treat one of the bodies as small.  
\citet{2011MNRAS.418.1043Q} consider three equal mass bodies, but on near-circular orbits with semi-major axes in a geometrical ratio.  
As mentioned above, other studies consider the possible formation of resonant chains due to the presence of a disk.  
\cite{Raymond2010} find that a planetesimal disk can easily produce 4:2:1 four-body resonances.  
Like \citet{Matsumura2010b}, \citet{Moeckel2012a} model planet-planet scattering during dissipation of a gas disk.  
The latter find that almost 45\% of their 3-planet simulations which remain stable achieve a 9:6:4, 6:3:2, 3:2:1, or 4:2:1 four-body resonance.  
Although no pair of planets in these simulations reduces to a 4:3 MMR, this prevalence of resonant chains suggests that a 4:3 MMR may be formed in this manner.

Another mechanism not directly considered in this paper is that of a chaotic migration scenario.
Here the continued inward migration of planets in a 3:2 MMR causes the eccentricity of the planets to increase to such an extent that the planets pass out of the portion of parameter space in which stable libration is possible, and into a region of chaotic scattering. 
In such a scenario, it is generally envisaged that without the protection afforded by being in a mean motion resonance, the planets will scatter one another during subsequent close approaches and the stable resonant configuration will be destroyed.
However, as the planets move into the region of chaotic scattering, an external perturbation (conceptually due to interactions with a disk or scattering from a small planetary embryo) might act to kick the planet from the chaotic overlap region down into the 4:3 MMR region.  
It is clear that in order for this scenario to realistically occur, the time-scale for a stochastic scattering event must be shorter than the time-scale over which chaotic scattering can start to disrupt the system. 
Detailed investigation would be required to understand whether such a scenario is at all possible, and more likely than the proposed scattering and damping scenario (see Section~\ref{sec:formation:scattering}).

\section{Small mass planets}\label{sec:smallmass}
\rev{
In this section, we extend the previous calculations and consider systems with small mass planets.
The motivation for this are the majority of systems in the Kepler sample which have a much smaller mass than those discovered by radial velocity (see Section~\ref{sec:observations:exo:kepler}). 
We will show that such systems can easily form via a variety of mechanisms and the problems discussed in the last section do not arise.
}

\rev{
We first look at the stability map of small mass planetary systems in Section~\ref{sec:small:stability:nbody}.
Then we go on and study the traditional migration-capture scenario with $N$-body simulations in Section~\ref{sec:small:formation:migration}.
In Section~\ref{sec:small:formation:embryo} we look at the possibility of growth of isolation mass embryos through direct collisions.
}

\subsection{Stability map with direct $N$-body simulations}\label{sec:small:stability:nbody}
\begin{figure}
\centering
\psfig{figure=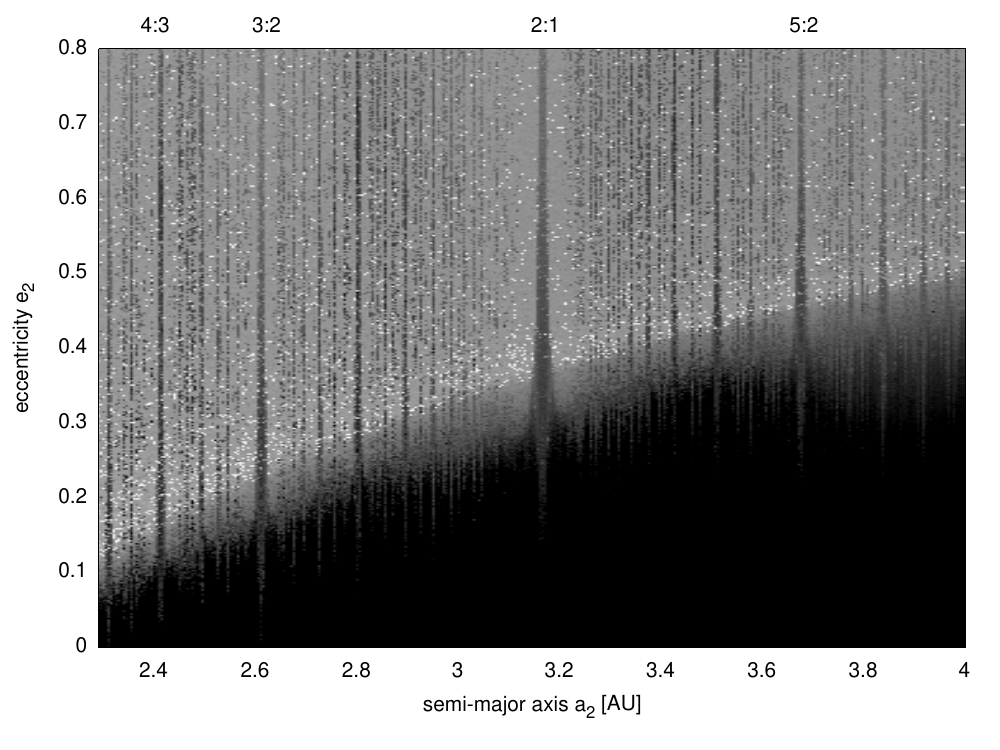,angle=0,width=1.0\columnwidth} 
\caption{Stability structure of the phase-space in a system with two planets.
The eccentricity and semi-major axis of the outer planet are varied.
The plotted quantity is the maximum Lyapunov exponent.
\rev{White regions are unstable, dark regions are stable.
This is the same plot as in Figure~\ref{fig:stability:nbody:widea2e2} but for small planetary masses, $m_1=m_2=\,\Mearth$.}
\label{fig:widea2e2earth}
}
\end{figure}

\rev{
The stability plots in Section~\ref{sec:stability:nbody} were dominated by large unstable regions and a small stable island associated with the 4:3 resonance. 
For two Earth mass planets, the stability plot looks very different, as shown in Figure~\ref{fig:widea2e2earth}.
Here, the parameter space of stable regions is much larger.
More structure is visible due to the presence of higher order resonances. 
In principle we can now capture into each of these resonance.
In practice, the capture probability depends of course on the migration rate.
We study this in the following section.
}

\subsection{Convergent migration in a disk}\label{sec:small:formation:migration}

\begin{figure}
\psfig{figure=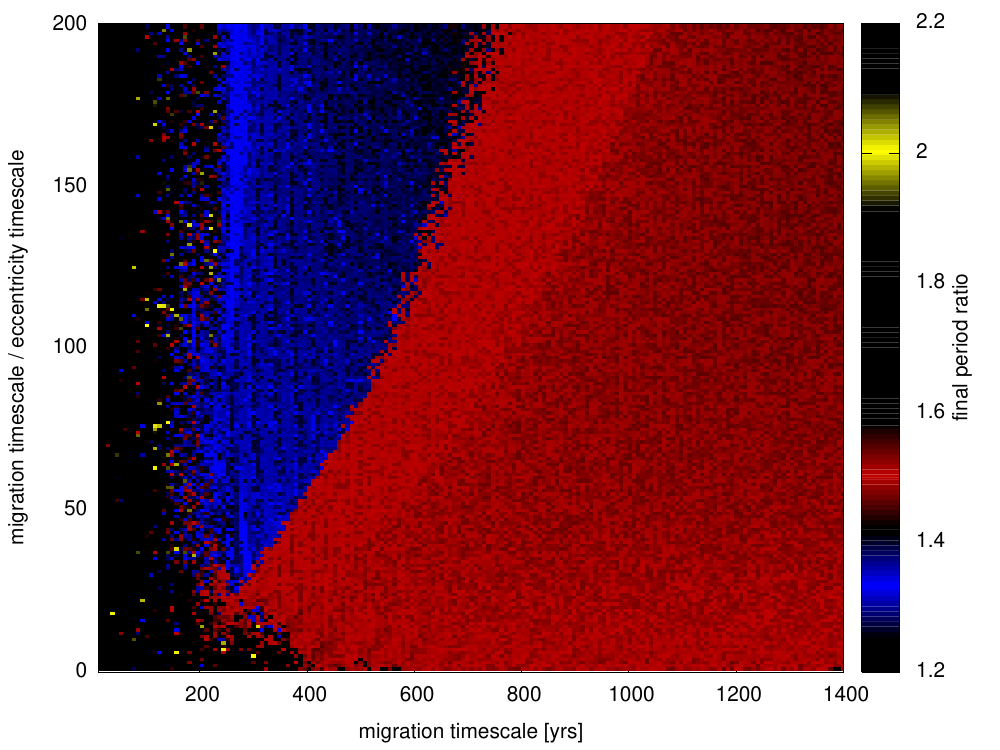,angle=0,width=1.0\columnwidth}
\caption{
Final period ratios after convergent migration of two planets as a function of the migration and eccentricity damping timescales.
Outer planet mass reduced compared to the system HD~200964: $m_1=1.8 \Mjup$, $m_2=0.09 \Mjup$.
\label{fig:formation:migration:nbody2}
}
\end{figure}

In \Figure{fig:formation:migration:nbody2} we plot the same quantity as in \Figure{fig:formation:migration:nbody1} but the mass of the outer planet has been reduced by a factor of ten. 
This leads to smaller critical migration rates required to pass through the 2:1 and 3:2~resonance. 
It also increases the stability of the system and thus enables us to form some stable systems which stay in the 4:3~resonance.
It is again helpful to recall the stability plots in Figures~\ref{fig:widea2e2} and~\ref{fig:widea2e2earth} to understand the differences in the evolutionary track.
\rev{The outer planet starts on the far right bottom of Figure~\ref{fig:widea2e2earth}. 
It migrates inwards (moving to the left) and encounters various commensurabilities.
Depending on the migration rate it captures into any of these resonances and gains eccentricity in the process (moves up). 
In contrast to the high mass example in Figure~\ref{fig:widea2e2}, the resonances are stable (dark color) and the planet does not have to cross an unstable part of the parameter space to get there.
}
Note that the migration rates required are still very high, $\tau_a \sim 200-800~\mathrm{yr}$. 
This corresponds to a type-III migration regime and is most likely not a realistic outcome of planet-disk interaction for these (reduced!) mass ratios.
This is the opposite to the reasoning of \cite{ReinPapaloizouKley2010}, where the mass ratios are such that type-III migration is considered to be the most likely formation scenario.

\subsection{Growth of isolation-mass embryos}\label{sec:small:formation:embryo}

The end stage of the runaway \citep{WetherillStewart1989,KokuboIda1996} and oligarchic growth \citep{IdaMakino1993,KI02} phases of planetary embryo growth is envisaged to be a series of isolation-mass embryos.
These are a series of embryos each having accreted all solid material within an annulus of width $c_\mathrm{iso}\,r_H$ of their location. 
Here, $c_\mathrm{iso}$ is a dimensionless constant of order $10$ and $r_H$ is the Hill radius of the embryo. 
For a power-law surface density profile $\propto \Sigma_0 a^{-\alpha}$, where $\alpha$ denotes the slope of the power law, the mass of the isolation mass embryos can be written as \citep[e.g. ][]{IdaLin2004}
\begin{eqnarray}
m_\mathrm{iso} &=& 5\cdot 10^{-3} c_\mathrm{iso}^{3/2} \left(\frac{\Sigma_0}{10\,\mathrm{g cm}^{-2}}\right)^{3/2}\nonumber\\
&&\quad\cdot\left(\frac{a}{1\mathrm{AU}}\right)^{2-\alpha} \left(\frac{M_{\star}}{M_{\odot}}\right)^{-1/2} M_{\oplus}.\label{eqn:iso}
\end{eqnarray}
We will use this as a starting point of our integrations.

\subsubsection{Methods to simulate the growth of isolation-mass embryos}\label{sec:small:formation:embryo:methods}
We follow a method conceptually similar to that employed in \citet{Zhou2007} and \citet{HansenMurray2011}.
We consider that a series of isolation-mass embryos has formed as described above.
We then simulate their subsequent evolution as they excite, scatter and collide over $\sim 10^9\,\mathrm{yr}$. 

The initial conditions used in Equation~\ref{eqn:iso} are set to be $\Sigma_0 = 10\,\mathrm{g\, cm}^{-2}$, $c_\mathrm{iso} = 7$ and $\alpha = 3/2$.
We distribute the embryos over annuli with widths either $0.55\,\mathrm{AU} < a < 1.75\,\mathrm{AU}$ (Set A) or $0.1\,\mathrm{AU} < a < 2.2\,\mathrm{AU}$ (Set B).

The embryos are damped for the first $10^6\,\mathrm{yr}$ of their evolution assuming an interaction with a gas disk of surface density of $\Sigma_0 = 2400\,\mathrm{g\,cm}^{-2}$ using the same eccentricity damping model as that employed in \citet{Mandell2007}. The gas disk is then allowed to dissipate on a time-scale, $\tau_d$, with this simply being modeled as a reduction in the damping force by a factor $e^{-t/\tau_d}$.
We use $\tau_d = 10^5\,\mathrm{yr}$, $10^6\,\mathrm{yr}$ and $10^7\,\mathrm{yr}$ for different subsets.

Typically, this results in the number of embryos $N$ remaining approximately constant for the first $10^6\,\mathrm{yr}$ as the strong damping prevents many crossing-orbits developing. 
In the case of Set A, $N$ is approximately $40$, in the case of Set B, $N \sim 130$.  As the damping begins to decrease (the gas disk dissipates), the embryos begin to excite and collide with one another, reducing the number of bodies present in the simulation.
The final number of planets varies across the simulations but is typically between 5 and 7.

\subsubsection{Results from $N$-body simulations of the growth of isolation-mass embryos}\label{sec:small:embryo:results}
\begin{figure}
\centering
\subfigure[The orbital periods as a function of time is shown in black.
As an illustration of the degree of eccentricity, the gray dashed lines show the the quantity $\left(a\,(1 \pm e)\right)^{3/2}$ for all of the bodies.  ]{
\includegraphics[width=0.95\columnwidth,trim= 2.5cm 1.5cm 2.5cm 2.5cm]{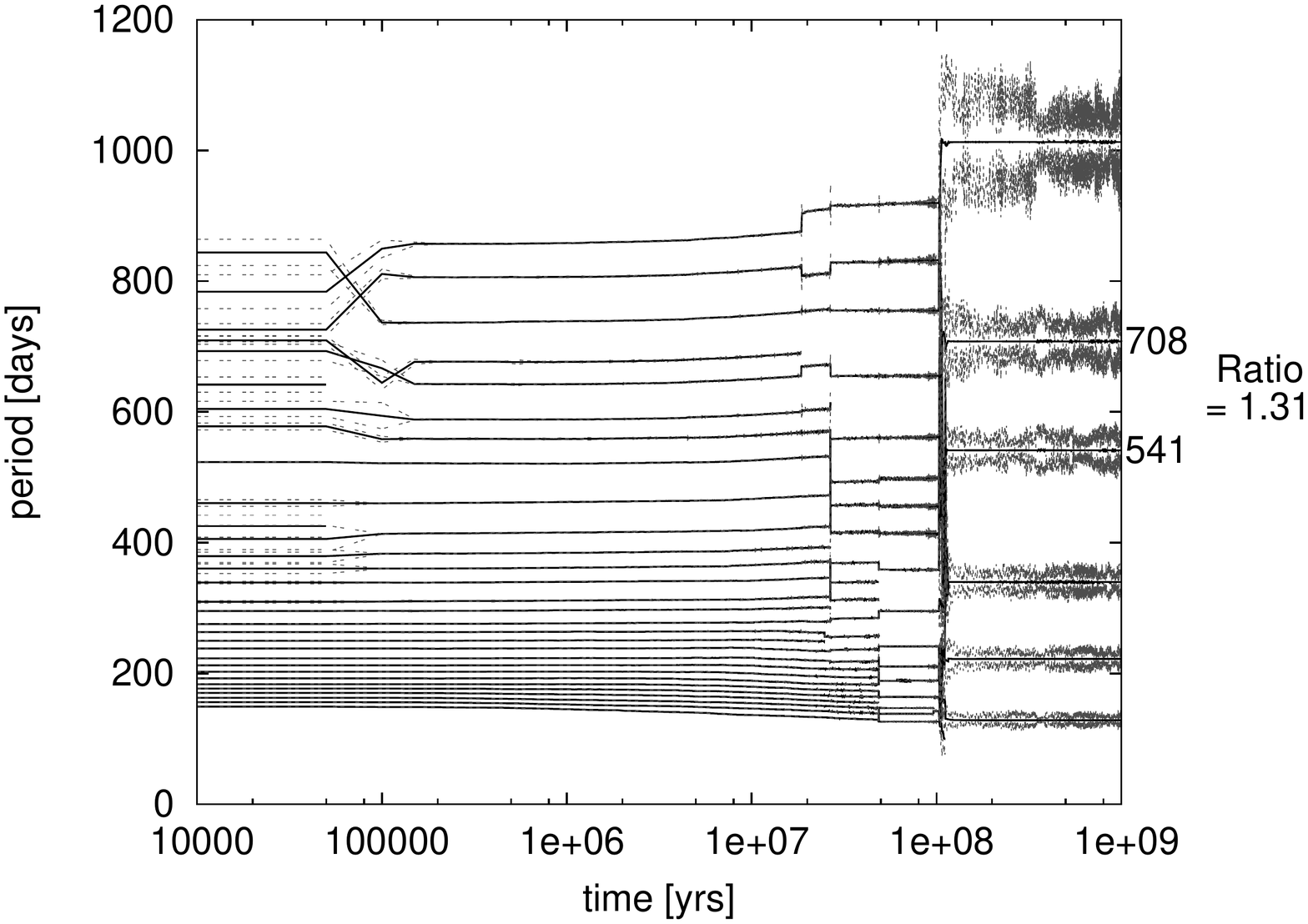}
\label{fig:formation:embryo:evolution1a}
}
\subfigure[The resonant angles $\theta_3,\,\theta_4$ and $\theta_{3-4}$ for the 4th and 5th bodies from the star are shown as a function of time. Only the secular resonance angle $\theta_{3-4}$ does librate for long periods.  ]{
\includegraphics[width=0.85\columnwidth,trim= 2.5cm 1.5cm 5.5cm 5.8cm]{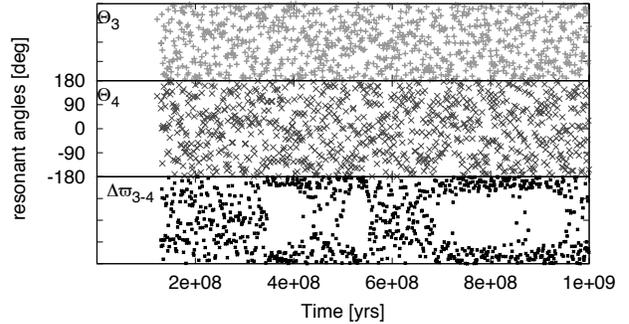}
\hspace{0.1\columnwidth}
\label{fig:formation:embryo:evolution1b}
}
\caption{Evolution of one damped embryo simulation. The damping time-scale is $\tau = 10^7\,\mathrm{yr}$. 
}
\label{fig:formation:embryo:evolution1}
\end{figure}

The results of a simulation run from Set A with $\tau  = 10^6 \mathrm{yr}$ are given in \Figure{fig:formation:embryo:evolution1}.
\Figure{fig:formation:embryo:evolution1a} shows the evolution of the periods as a function of time. 
At $\sim 10^9\,\mathrm{yr}$ the 4th and 5th planets from the star (highlighted in the example) have period ratios $\sim 1.33$, i.e. close to the 4:3 MMR period ratio. 
In \Figure{fig:formation:embryo:evolution1b} we show the resonant angles $\theta_3,\,\theta_4$ and the secular resonance angle $\theta_{3-4}$ defined in Equation~\ref{EQN:RES:ANGLES} for the 4th and 5th bodies from the star.
The resonant angles are observed to circulate over the full $360^{\circ}$ range, while the secular angle librates for extended periods (although over very long time periods it too can be seen to circulate). 
To understand in detail the strengths of these resonances and the fraction of time that the planets spend in (or adjacent to) these MMRs as a function of (e.g.) the overlap of 3-body resonances requires much more detailed analysis in the manner of \citet{2011MNRAS.418.1043Q}. 

\begin{figure}
\centering
\begin{tabular}{c}
\includegraphics[width=0.95\columnwidth,trim= 2.5cm 1.5cm 2.5cm 2.5cm]{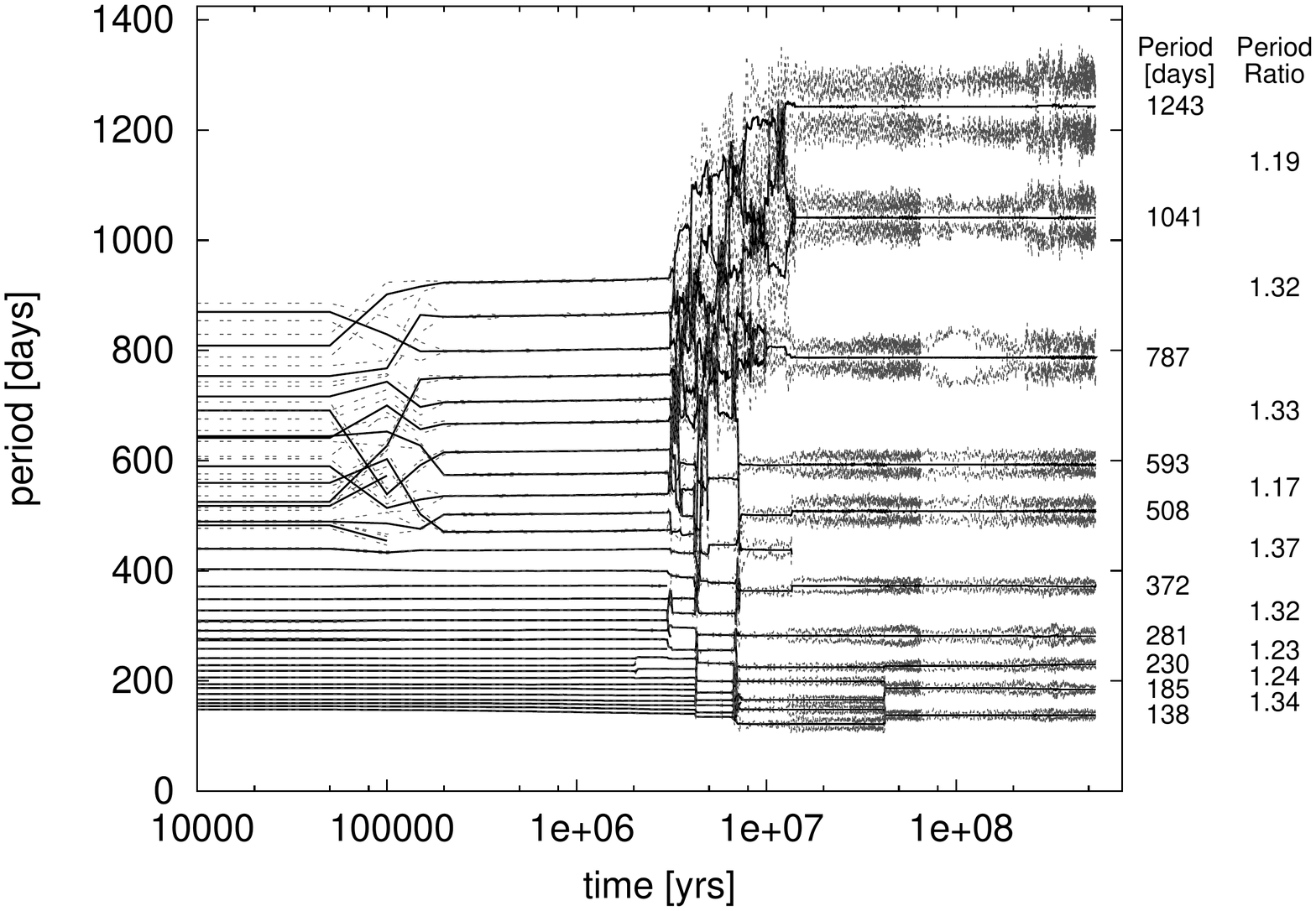}\\
\end{tabular}
\caption{Evolution of one damped embryo simulation.
The damping time-scale is $\tau = 10^6\,\mathrm{yr}$. 
The orbital periods as a function of time is shown in black.
As an illustration of the degree of eccentricity, the gray dashed lines show the the quantity $\left(a\,(1 \pm e)\right)^{3/2}$ for all of the bodies. 
The number of embryos reduces from $\sim 40$ to $10$, and a number of very close period ratios are present between various pairs of the planets.
}
\label{fig:formation:embryo:evolution2}
\end{figure}

It is also of interest to note that in the simulations we performed, more closely spaced planetary configuration also arose.
In \Figure{fig:formation:embryo:evolution2} we illustrate the results of a simulation run from Set A with $\tau = 10^7\mathrm{yr}$. 
In this system ten low mass bodies survive at $t\sim 5\times 10^8\,\mathrm{yr}$.
These have period ratios $1.34,1.24,1.23,1.32,1.37,1.17,1.33,1.32$ and $1.19$.
Hence multiple pairs are close to the 4:3 (1.33), 5:4 (1.25) and 7:6 (1.17) MMR period ratios. 
We have checked (but do not plot) the various possible two-body resonant angles for the respective pairs of planets close to the listed MMRs. 
We find that none of the bodies occupy exact 2-body resonances despite their period ratios being suggestively close. 

We thus find that the formation of rather closely spaced systems (i.e. with period ratios $1.15\rightarrow 1.4$) is easily accomplished through the collisional evolution of isolation-mass embryos in an extended catastrophic collision phase of planet formation.
But we emphasize that period commensurabilities do not equate to two-body resonances.

The total embryo mass considered in these simulations was low.
For Set A we have $\sim 3.0 M_{\oplus}$ and for Set B $\sim 6.0 M_{\oplus}$.
This corresponds to solid surface densities consistent with a minimum mass nebular model.
The absolute masses of the planets formed at $t\sim 10^9\,\mathrm{yr}$ are subsequently low ($0.5 M_{\oplus}$ and $0.55  M_{\oplus}$ for the example in \Figure{fig:formation:embryo:evolution1}, $0.1 - 0.54M_{\oplus}$ for the bodies in \Figure{fig:formation:embryo:evolution2}). 
While these masses are likely smaller than those listed in \Table{tab:KOI}, they are still within the range of detectable masses for Kepler \citep[e.g. KOI-961, see][]{Muirhead2012}. 
All of the systems that we were able to form have multiple planets.
We also note that \citet{HansenMurray2011} found nothing conceptually different in their simulations performed at much higher absolute surface density normalizations.
These simulations resulted in mass-period distributions approximately resembling those of the observed Kepler systems.

We emphasize that these simulations have run for $t \sim 10^9\,\mathrm{yr}$.
For the majority of this time period the planets have been undamped.
It is therefore realistic to consider these systems as being in an old, evolved state as might be observable by the (e.g.) Kepler mission. 
This mechanism may therefore explain the KOIs listed in \Table{tab:KOI}. 

The growth of planets from isolation-mass embryos seems to provide a natural means by which low mass planets can be either captured into closely-spaced MMRs (i.e. 4:3), or have period commensurabilities approximately similar to such closely-spaced MMRs while also remaining long-term stable.
We emphasize that while such close spacing (period ratios less than $1.4$) does not happen in the majority of systems simulated, it was common enough for us to observe it in 2 out of 60 systems simulated.

\section{Conclusion}\label{sec:conclusions}
In this paper, we explored numerous methods of forming a 4:3 mean motion resonance in systems of both \rev{high-mass planets}.
We found that it is extremely difficult to form stable massive systems in a 4:3 resonance. 
The discovery of multiple such systems in radial velocity surveys leads us to conclude that traditional formation methods are failing to reproduce the observed fraction of systems that are in or close to this resonances. 

More precisely, we found that convergent migration due to torques imposed by a gas disk is not a viable mechanism for the formation of massive planets in a 4:3 resonances. 
In Section~\ref{sec:formation:migration} we showed that it can be ruled out to form the system HD~200964 in this way.
There are four reason for this. First, the systems tend to lock into higher order resonances. Second, even with reduced masses the required migration timescale to pass through resonances such as 2:1 and 3:2 is unphysically short. Third, gap opening tends to stall migration at gap edges which prevents subsequent capture into close-in resonances. Fourth, the surfing effect pushes planets away from each other \rev{unless both planets are able to open a clean gap}.
We conducted a large survey of hydro-dynamical simulations. 
But we did not find a way to realistically form a 4:3 resonance in any of these hydro-dynamic models.

It is important to note that these are planets detected by the radial velocity method and thus the masses are only minimum masses. 
However, we found that masses higher then the minimum masses make it even harder to form a stable resonance.

In-situ formation where planet embryos start out in the 4:3 resonance is also not a viable formation mechanism. 
In Section~\ref{sec:formation:insitu} we conducted hydro-dynamical simulations of such a scenario.
There is always a phase when the planets migrate divergently during their mass accretion phase.
As soon as the planets move apart from each other, outside of the 4:3 resonance, they cannot recapture into the resonance at a later stage because they preferentially capture in wider resonances.
Gap opening and the tidal surfing effect are further preventing planets from staying or recapturing in a tight resonance while being embedded in a gas disk.

A combined scattering and damping mechanism does seem to be a plausible means of forming giant planets in closely-spaced MMRs as we showed in Section~\ref{sec:formation:scattering}.
We found that for a wide range of damping parameters close-in resonance can form. 
However, the initial conditions are finely tuned to allow such a capture. 
We estimate that the fraction of planetary systems that might end up in such a configuration is much smaller than the currently observed fraction of approximately $4\%$.
This scenario also predicts the formation a large number of planets in a 1:1 co-orbital resonance which is not consistent with current observations.

\rev{We extended the study to include small mass planets in foresight of several Kepler planet candidates which are close to a 4:3 period ratio. 
We found that there is no problem in forming any of these smaller mass planets with a traditional migration scenario. 
We therefore expect that several of these systems will be confirmed by follow up observations.}

\rev{The main result in our paper is a negative one, i.e. we cannot explain the formation of the observed systems.
One could use our results as evidence against the existence of massive planets in close in resonances and search for other explanations of the observed RV signals. 
But this requires a great amount of caution. 
Using the same argument, we could have ruled out the first discovered Hot Jupiter. 
Rather, we hope that this study will guide future investigations and spur interest in these systems among the community. 
}

\section*{Acknowledgments}
\rev{We thank Alessandro Morbidelli for an extremely helpful and detailed referee report that greatly improved this paper.}
Hanno Rein would like to thank Scott Tremaine, Alice Quillen, Eiichiro Kokubo and Ilona Ruhl for helpful comments at various stages of the project.
Eric Ford and Matthew Payne acknowledge enlightening discussions with Althea Moorhead and Richard Ruth,
and \rev{Dimitri Veras thanks Alex Mustill for helpful comments}.  Hanno Rein was supported by the Institute for Advanced Study and the NSF grant AST-0807444.
Matthew Payne was supported by NSF grant AST-0707203 and the NASA Origins of solar systems grant NNX09AB35G.

\appendix

\section{Additional stability plots}\label{app:widea1a2}

\begin{figure}
\centering
\subfigure[Small eccentricities $e_1=e_2=0.1$.]{
\psfig{figure=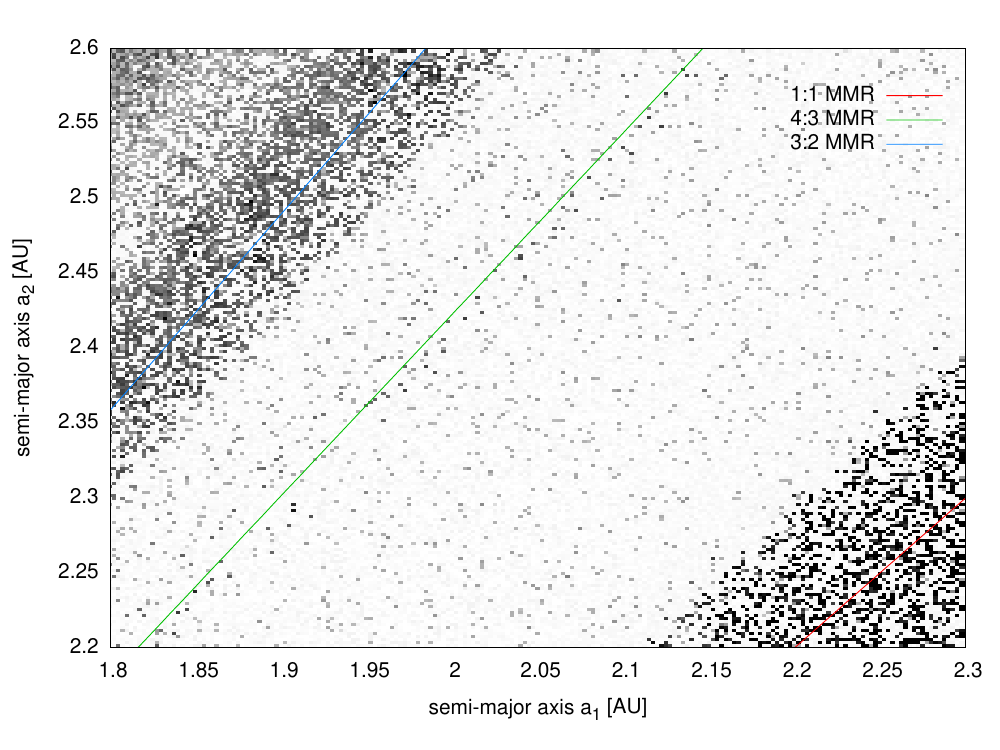,angle=0,width=.960\columnwidth} }
\subfigure[High eccentricities $e_1=e_2=0.5$.]{
\psfig{figure=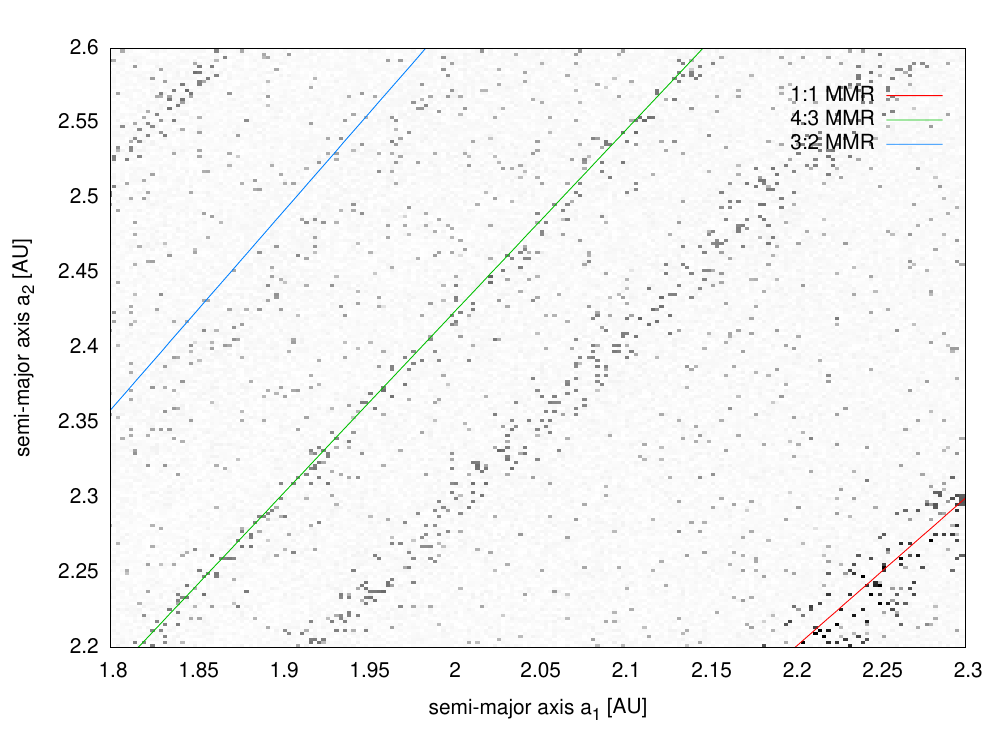,angle=0,width=.96\columnwidth} }
\caption{Stability structure of the phase-space for two planets in the $a_1,a_2$ plane.
\rev{White regions are unstable, dark regions are stable.}
}
\label{appfig:widea1a2}
\vspace{.5cm}
\psfig{figure=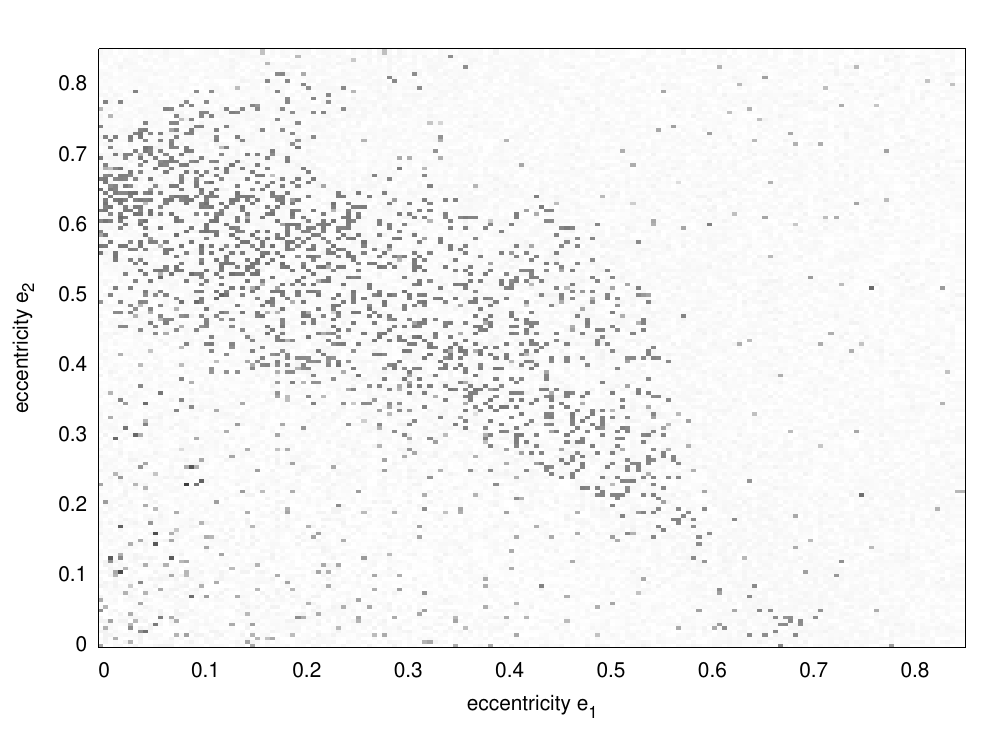,angle=0,width=.96\columnwidth} 
\caption{Stability structure of the phase-space for two planets in the $e_1, e_2$ plane.
\rev{White regions are unstable, dark regions are stable.}
}
\label{appfig:widee1e2}
\end{figure}

In this appendix we show additional stability plots of the phase space of two massive planets.
The procedure is described in Section~\ref{sec:stability:nbody}. 

\Figure{appfig:widea1a2} shows two additional stability plots of the phase space in the $a_1$-$a_2$ plane. 
The initial eccentricities are $0.1$ and $0.5$ for the top and bottom plot, respectively. 
All other angles are drawn from a uniform distribution. 
The system is assumed to be coplanar.
One can see that there is clear evidence of resonances as the stability only depends on the period ratio (not the individual periods) and the eccentricities of the planets.

\Figure{appfig:widee1e2} shows an additional stability plot of the phase space in the $e_1$-$e_2$ plane. 
The planets' semi-major axes are initially at $a_1=2$ and $a_2=2.42$, i.e. close to a 4:3~period ratio. 
We find stable islands for a wide variety of eccentricities as long as $e_1<0.55$.
One can see a clear anti-correlation of the eccentricity of the inner and outer planet.

\section{Forced migration for inclined planets}\label{app:nbody:inclined}
\begin{figure}
\centering
\psfig{figure=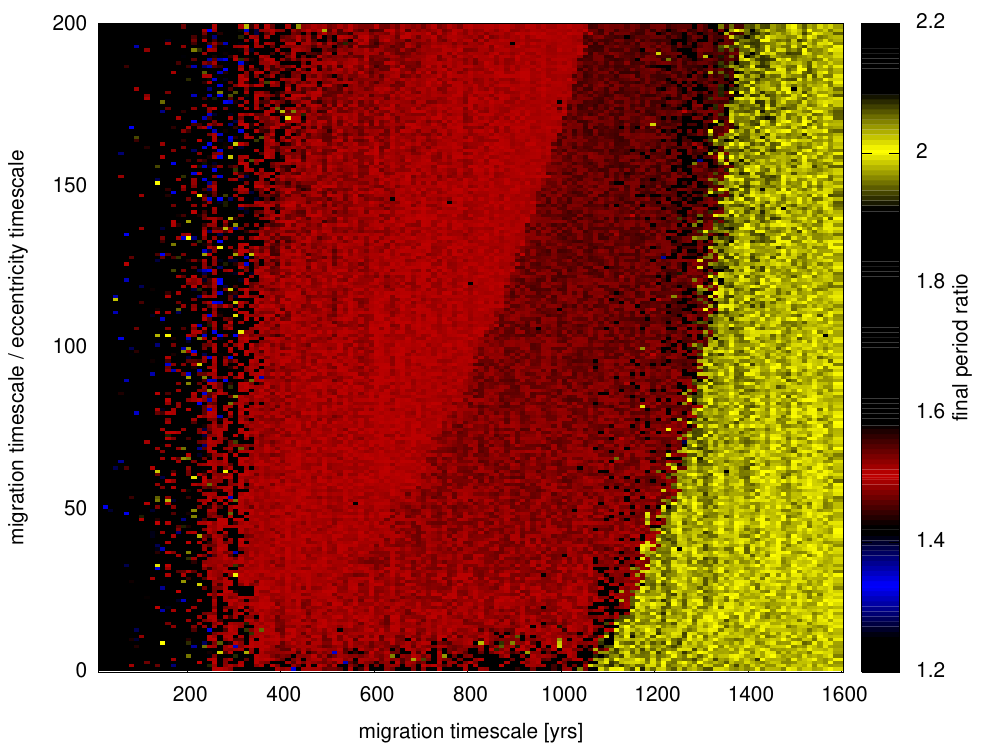,angle=0,width=1.0\columnwidth }
\caption{Final period ratio of two convergently migrating planets with an initial mutual inclination of $i=5^\circ$. }
\label{appfig:inclined}
\end{figure}

We show the final period ratio of two convergently migrating planets in \Figure{appfig:inclined}.
The simulations are identical to those presented in \Figure{fig:formation:migration:nbody1} but are fully three dimensional and include a finite initial inclination of $i=5^\circ$ between the planets. 

There is no inclination damping present in this simulation.
However, in contrast to \cite{TremaineYu2000}, we include explicit eccentricity damping which prohibits substantial eccentricity and inclination growth.

One can see no qualitative or quantitative change in the results compared to the coplanar case. 
This allows us to restrict ourselves to the coplanar formation scenarios presented above which reduces the number of free parameters in the initial conditions.

\section{Illustration of Capture into 1:1 Resonance}\label{app:onetoone}
\begin{figure}
\centering
\begin{tabular}{c}
\includegraphics[width=0.9\columnwidth,trim= 2.5cm 1.5cm 2.5cm 2.5cm]{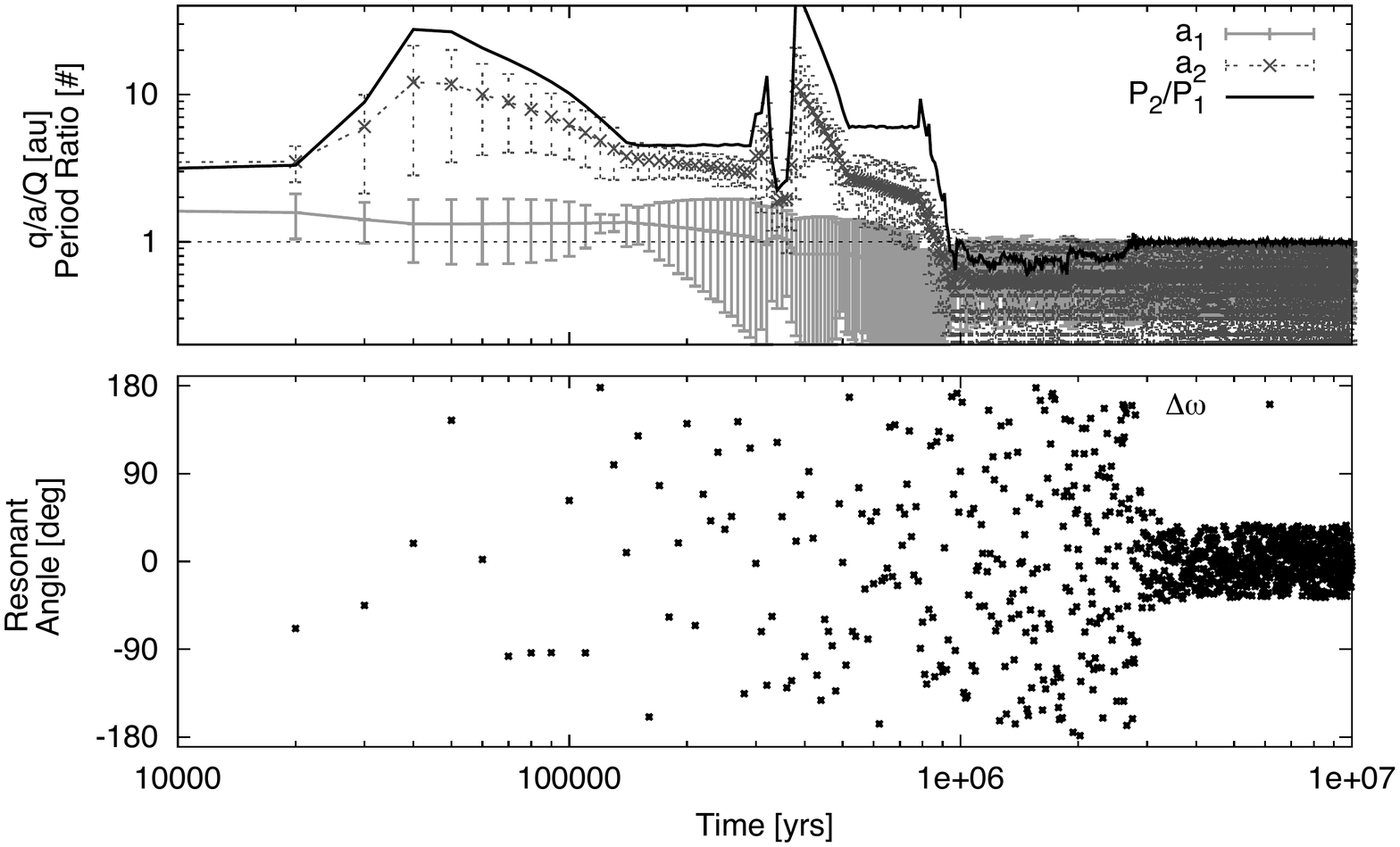}\\
\end{tabular}
\caption{
Capture into 1:1 resonance after a scattering event. 
\textit{Top:} This plot shows the semi-major axes (in AU, bottom two curves) and the period ratio (top black curve) as a function of time. 
\textit{Bottom:} This plot shows the resonant angle $\omega_2 - \omega_1$ of the two planets as a function of time.
}
\label{fig:onetoone:evolution1}
\end{figure}

In \Figure{fig:onetoone:evolution1} we plot an example of the capture of two giant planets into a resonance 1:1 resonance. 
The simulation methodology is exactly as described in Section~\ref{sec:small:formation:embryo:methods}.
The initial conditions for the inner planet are
$m_1 = 1.8 \Mjup$, $a_1 = 1.6 \mathrm{AU}$ and $e_1 = 0$, and for the outer planet
$m_2 = 0.9 \Mjup$, $a_2 = 3.6 \mathrm{AU}$ and $e_2 = 0.46$ ($q_2 = 1.92$).

The damping parameters are the same as those applied for \Figure{fig:formation:embryo:evolution1}. The migration timescale is $\tau = 10^5\,\mathrm{yr}$ and $K = 1$.
The disk removal timescale is $\tau_d=10^7\,\mathrm{yr}$.
The damping is operating on the outer body only, for all $r > 1 \mathrm{AU}$.

The outer planet migrates inwards over the first $10^6\,\mathrm{yr}$ of the simulation, while simultaneously suffering occasional strong perturbations from the inner planet. 
At $t \sim 6\times 10^5\,\mathrm{yr}$, the pericenter of the outer planet drifts inside the apocenter of the inner planet and a period of strong interaction and rapid inward migration commences, leading to capture into the 1:1 resonance.
The resonance remains stable until the end of the simulation.

\bibliographystyle{aa} 
\bibliography{full}

\label{lastpage}
\end{document}